\documentclass[a4paper,12pt]{article}
\usepackage{setspace}
\usepackage[english]{babel}
\usepackage[latin1]{inputenc}
\usepackage[pdftex]{graphicx}
\usepackage{theorem}
\usepackage[french]{varioref}
\usepackage{amsmath}
\usepackage{ifthen}
\usepackage{commath}
\usepackage{amsfonts}
\usepackage{float}
\usepackage{color}
\usepackage{bibentry}
\usepackage[toc,page]{appendix} 
\usepackage{multirow}
\usepackage{hyperref}
\nobibliography*

\newtheorem{prop}{Proposition}

\newcommand{\lat}{\mathrm{Lat}}
\newcommand{\lgtd}{\mathrm{Long}}

\newcommand{\M}{\mathrm{M}}
\newcommand{\N}{\mathrm{N}}
\newcommand{\R}{\mathrm{R}}
\newcommand{\cov}{\mathrm{cov}}
\newcommand{\var}{\mathrm{var}}
\newcommand{\E}{\mathrm{E}}

\begin{document}

\title{Gaussian linear state-space model for wind fields in the North-East Atlantic}

\author{Julie Bessac\textsuperscript{1}, Pierre Ailliot\textsuperscript{2}, Val\'erie Monbet \textsuperscript{1}}
\date{}
\vspace{20mm}
\maketitle

{\small $^1$ \em Institut de Recherche Math\'ematiques de Rennes, UMR 6625, Universit\'e de Rennes 1, France}\\
{\small $^2$ \em Laboratoire de Math\'ematiques de Bretagne Atlantique, UMR 6205, Universit\'e de Brest, France}\\

\begin{abstract}

A space-time model for wind fields is proposed. It aims at simulating realistic wind conditions with a focus on reproducing the space-time motions of the meteorological systems. A Gaussian linear state-space model is used where the latent state may be interpreted as regional wind condition and the observation equation links regional and local scales. Parameter estimation is 
performed by combining a method of moment and the EM algorithm whose performances are discussed using simulation studies. The model is fitted to 6-hourly reanalysis data in the North-East Atlantic. It is shown that the fitted model is interpretable and provide a good description of important properties of the space-time covariance function of the data, such as the non full-symmetry induced by prevailing flows in this area.

\vspace{.5cm}

 
\end{abstract}


\section{Introduction}\label{section1}

Many natural phenomena and human activities depend on wind conditions. Meteorological data are often available over periods of time that are not long enough to estimate reliably probabilities of complex events. Stochastic weather generators have been developed to overcome this insufficiency by simulating  sequences of meteorological variables with statistical properties similar to the ones of the observations.  They have been adopted in impact studies
as a computationally inexpensive tool. Wind generators have  in particular  been used to assess wind power production (see \cite{BRO84,HOF13}), drift of objects in the ocean (see \cite{AILL06drift}) or coastal erosion (see \cite{SKID90}).

A review of stochastic models simulating artificial wind time series can be found in \cite{MONB07}. The most classical approach for modeling wind time series at a single location consists in using the Box-Jenkins methodology, where an ARIMA model is fitted after applying a marginal transformation to obtain Gaussian like  margins. The most usual transformation is a power transformation  (see \cite{BRO84}, \cite{RAFT89}, \cite{NFA96}, \cite{KAM97}), but specific distributions are  used as well, for instance Weibull \cite{BRO84}, truncated Gaussian \cite{GNEI06} or skew distributions \cite{HER10}. The conditional mean (and  variance) is then modeled given the wind at the previous time step. Non-linear models have also been proposed (see \cite{AILL12} and references therein). 

Generalizations to space-time models have been explored more recently. Multisite wind models have to deal with temporal and spatial dependence and it is known that these two components  are not separable \cite{GNEI02}. Black box models like artificial neural network can be fitted but they lead to non interpretable models. Two other approaches have been detailed in the literature: the models based on Gaussian fields (or Gaussian vector) with a parametric non separable covariance function \cite{RAFT89, RYC13} and the models based on Vector Autoregressive models (VAR) \cite{LUN05}. Both approaches allow to characterize space and time variability of the wind and, in particular, the motions of the air masses. In the Gaussian field method, these displacements are characterized by the interaction of time and space in the covariance function (\cite{RYC13}, \cite{RAFT89}, \cite{GNEI02}). One difficulty of  these models is to infer the parametric form of the covariance function and standard models  are generally used such as Gaussian or Matt\'ern covariances.  In VAR models, motions are introduced using covariates or switchings.   For example, in \cite{GNEI06}, regimes describing the main weather types (westerly/easterly wind) are introduced and different VAR models are fitted in each regime. In \cite{HER10} the wind direction is introduced in the VAR model.  In \cite{AILL06}, the VAR coefficients depend on a latent process which describes the motion of the air masses. In \cite{SAL11}, a latent field  describes the spatial structure of the AR parameters.

In VAR models two scales are implicitly modeled: a regional scale representing the prevailing flows and a local one corresponding to the locally observed wind conditions. But  VAR models may lead to over parametrization, especially when sites are highly correlated. In the present paper, a new approach is investigated. The regional wind is explicitly introduced as a latent variable, with its own autoregressive dynamic, and the local wind is expressed as a function of the regional wind at different lags to model the mean displacement of the air masses. The model is kept simple and interpretable since it is a linear Gaussian state-space model. Statistical inference can thus be performed efficiently and covariates can  easily be added. Despite of its simplicity, the model leads to non separable and anisotropic covariance functions.

The data considered in this paper are presented in Section~\ref{data}. The model is described in Section~\ref{model}. Parameter estimation and fitting procedures are also discussed in this section. Validation of the model si discussed in Section~\ref{results}. It is shown that the fitted model is able to reproduce the anisotropy and non-separability of the data. Various reduced models are introduced in Section~\ref{parsimonious model} and  conclusions are given in Section~\ref{discussion}. Parameter identifiability and non full-symmetry are proven in Appendix~\ref{nonidentif}.


\section{The wind dataset}\label{data}

In situ data are neither available on a long time period nor on a large area offshore Brittany in France. For a reliable study we choose to use reanalysis data which are obtained by combining observations with numerical weather prediction models. It provides relevant datasets for meteorological or climatological studies.
The data under study are wind speed intensities at $10$ meters above sea level extracted from the ERA Interim Full dataset produced by the European Center of Medium-range Weather Forecast (ECMWF). It can be freely downloaded and used for scientific purposes at the URL \emph{http://data.ecmwf.int/data/}.

\begin{figure}
\begin{center}
\includegraphics[width=6.5cm]{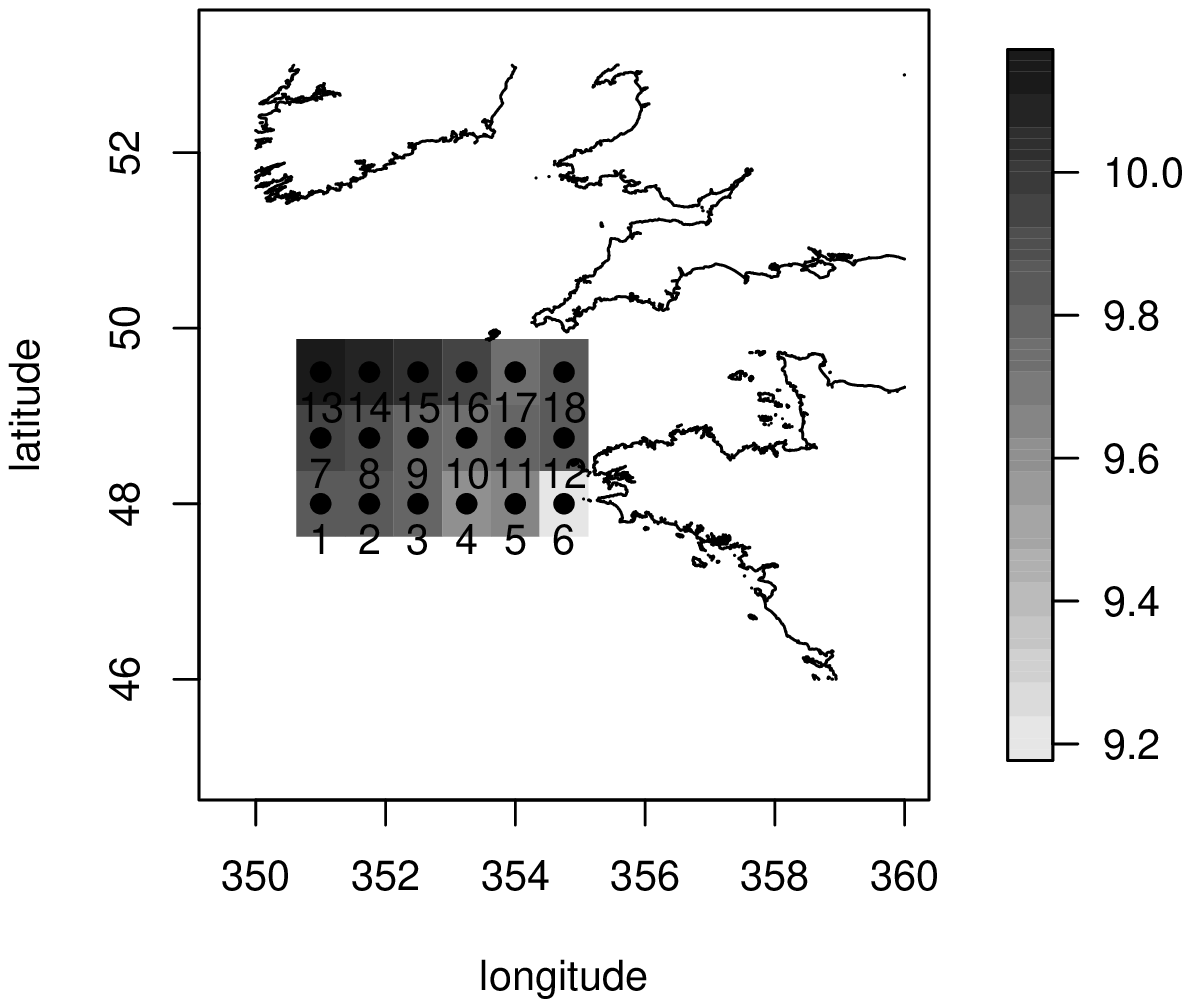}
\includegraphics[width=6.5cm]{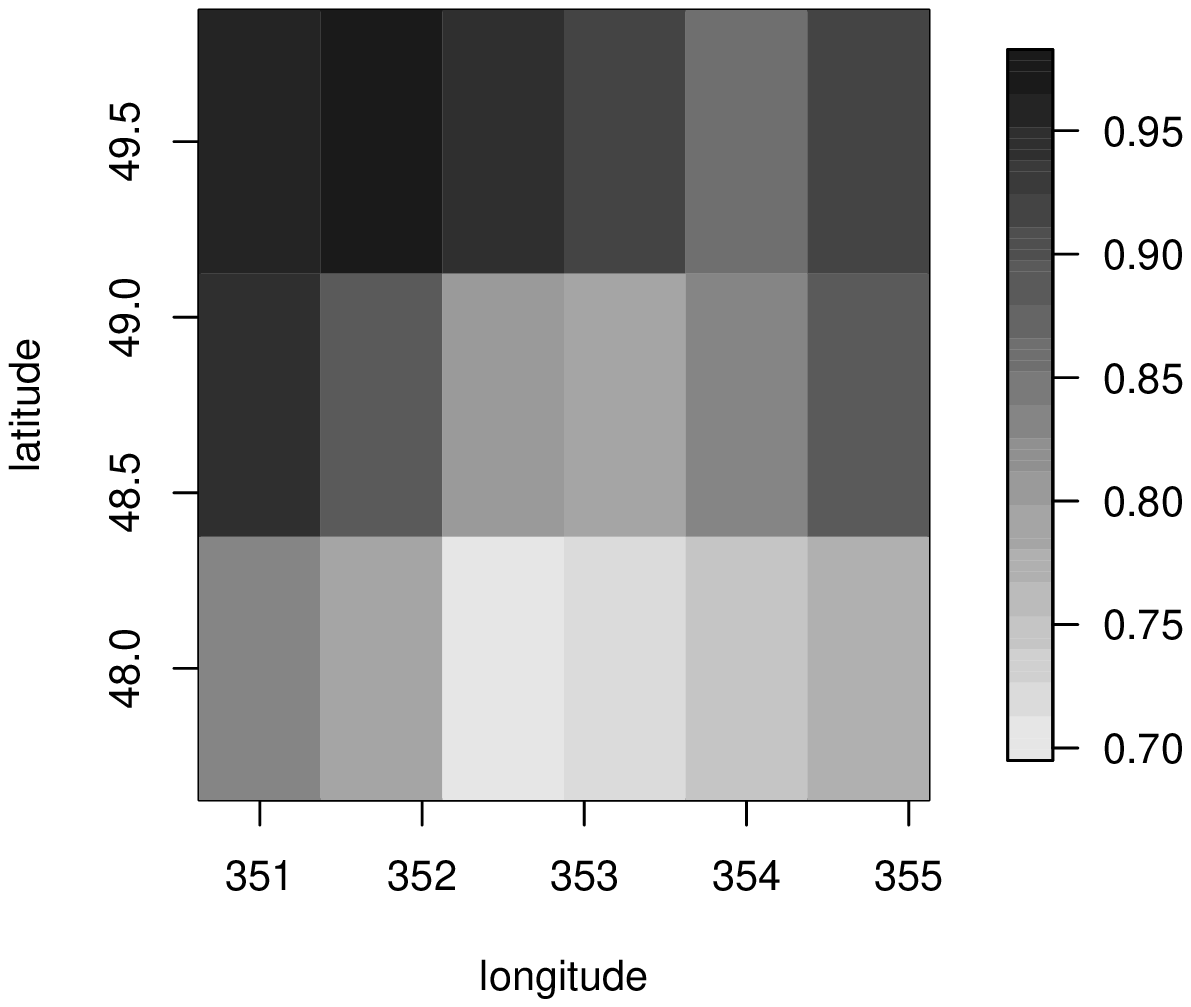}
\caption{Left panel: mean wind speed at the 18 numbered points under study in the North-East Atlantic. Right panel: estimated values of the power in the Box-Cox method at the 18 locations.}
\label{map}
\end{center}
\end{figure} 

This dataset is available on a regular space-time grid with a temporal resolution of $6$ hours and a spatial resolution of $0.75^{\circ}$. The methodology introduced in this paper could however easily be adapted to handle datasets with more complicated space-time sampling such as the one obtained when considering networks of meteorological stations. We focus on $18$ gridded locations between latitudes  $48^{\circ}$N and $49.5^{\circ}$N and longitudes $6.25^{\circ}$W and $9^{\circ}$W (see Figure \ref{map}).  The dataset consists of 33 years of wind data from $1979$ to $2011$ and we focus on the month of January. Further, the statistical inference is based on the assumption that the 33 months of January wind data are 33 independent realizations of a common stochastic process. This assumption is not unusual for meteorological processes but it does not take into account low frequency variations such as the the North Atlantic Oscillation (NAO).

\begin{figure}
\begin{center}
\includegraphics[width=6.5cm]{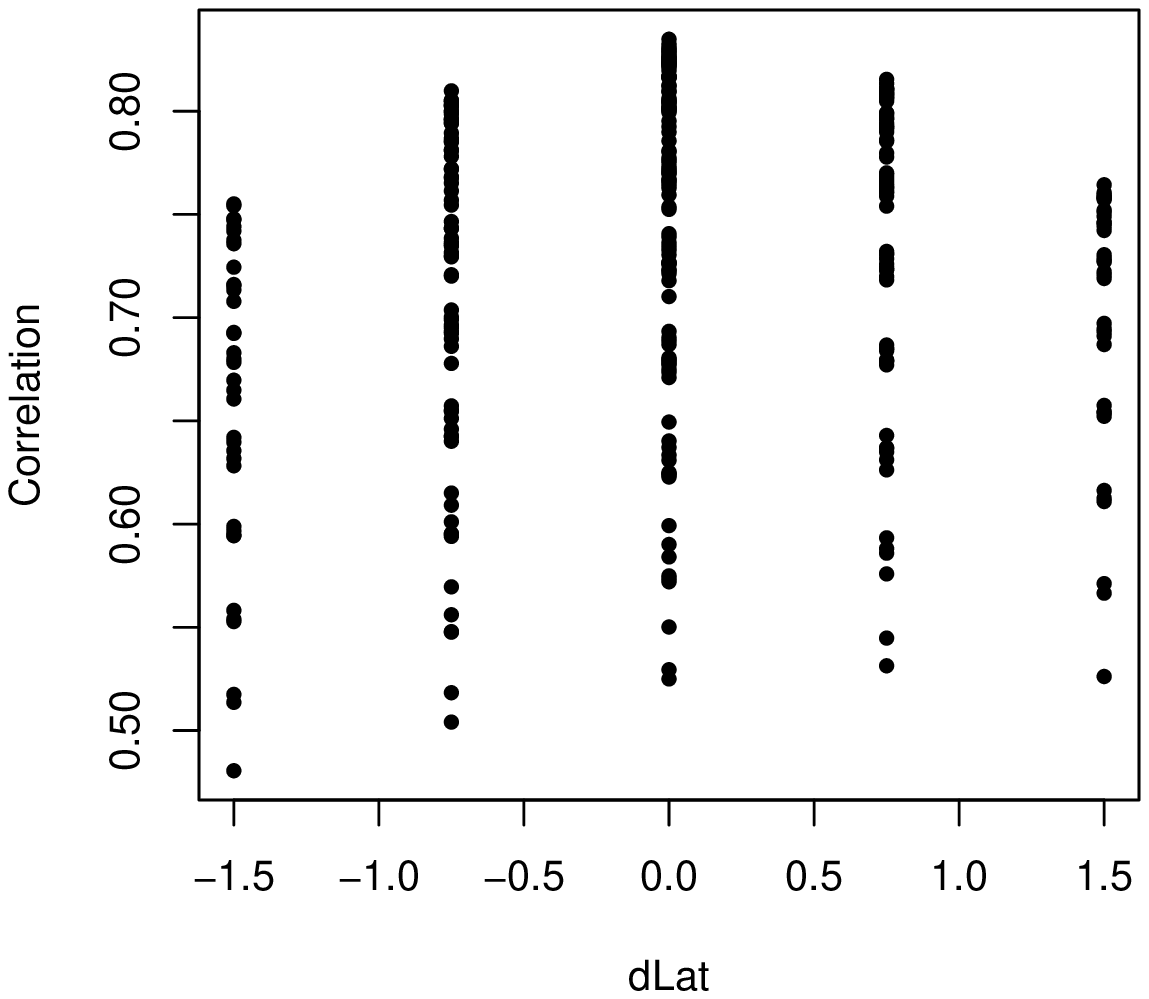}
\hglue .1cm 
\includegraphics[width=6.5cm,angle=90]{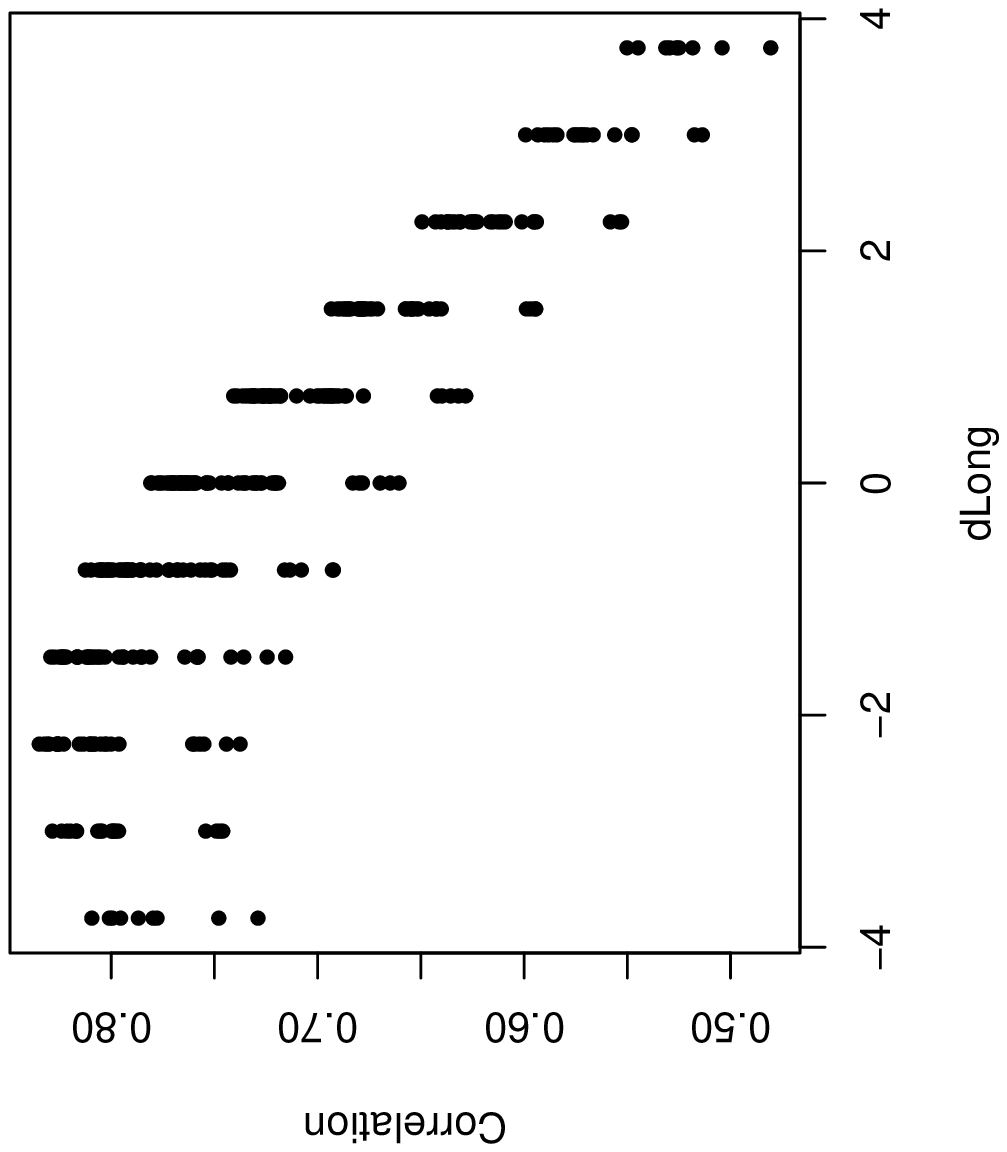}
\caption{Lagged one cross-correlations against differences of latitude (left) and longitude (right).}
\label{corrVSlatlong}
\end{center}
\end{figure}

In the studied area prevailing air masses are generally moving eastward. It creates non-separability and non full-symmetry properties of the associated space-time covariance function (see \cite{GNEI02}). In Figure \ref{corrVSlatlong} lagged one cross-correlations  highlight this phenomenon. The asymmetry with respect to the difference of longitude reveals that the correlation between $Y_{t}(p)$ and $Y_{t+1}(p')$ is higher when location $p$ is more westerly than $p'$ than when $p$ is easterly with respect to $p'$. This asymmetry is less pronounced in latitude but reveals flows from north to south. Full-symmetry is rejected as for the famous dataset of wind speed in Ireland studied in \cite{RAFT89}, \cite{GNEI02}, \cite{LUN05} and \cite[chapter 4]{GNEI07}. As in Figure \ref{corrVSlatlong} the study of correlations at lag 0 reveals some anisotropy as dependences in latitude and longitude differ.

Wind speed distribution is known to be skewed. It is generally modeled as a Weibull distribution (see e.g. \cite{BRO84}) but other distributions such as the skew normal distribution have also been considered (see \cite{FLEC10}). A classical method to handle such asymmetry in time series analysis consists in applying a Box-Cox transformation in order to get a time series with approximately marginal Gaussian distribution and in fitting an ARMA model to the transformed time series. This method has been extensively used for analyzing wind time series (see e.g. \cite{BRO84}, \cite{RAFT89}, \cite{NFA96}, \cite{KAM97}) and is generalized in a space-time context here.
%
%
%
%
It has been proposed to use a different transformation at each location (see \cite{RYC13}) but we have chosen to use the same power transformation at all sites in order to preserve the spatial structure of the wind fields following e.g. \cite{RAFT89}. More precisely, we consider
$$\left \{\begin{array}{lcl}
   y_{\lambda,i,t} & = &  \frac{y_{i,t}^{\lambda}-1}{\lambda} \quad \textrm{if $\lambda>0$}\\
   y_{\lambda,i,t}  & = & \log (y_{i,t}) \quad\textrm{ if $\lambda=0$}. \\
  \end{array}
\right.$$
with $y_{i,t}$ the wind speed at time $t$ and location $i$. The value $\hat \lambda = 0.85$ has been used in the sequel.  It is derived as the mean value of the $\hat \lambda_{i}$ depicted on Figure \ref{map} which are obtained at individual locations with the following criterion given in \cite{HIN77}. It is based on the empirical measure of asymmetry
$$S(\lambda_{i})= \frac{\textrm{mean}(y_{\lambda_{i},i,t} ) - \textrm{median}(y_{\lambda_{i},i,t} )}{\sqrt{\var(y_{\lambda_{i},i,t} )}}$$
and $\lambda_{i}$ is chosen as equaling to $0$ this statistic. Figure 1 shows that the data close to the coast are less Gaussian than the offshore ones. The model will be fitted on this power transformed and mean corrected data. For sake of simplicity we denote $Y_t$ the transformed observed field instead of $Y_{\lambda,t}$ in the sequel. \\



\section{A linear Gaussian state-space model for wind speed}\label{model}

State-space models first appeared in engineering and have then been extensively used in many domains. State-space representations bring a very flexible framework for modeling time series (see \cite{DUR01} and \cite{BROC06}) and space-time processes (see \cite{WIK10}). The model introduced in this Section is a linear Gaussian state-space model.  One of the main advantages of this class of model estimation, is that forecasting and smoothing can be processed trough general and efficient procedures. 





\subsection{Model}

The observed wind fields are generally smooth leading to a high correlation between the different sites. It suggests that it may be possible to explain an important part of the signal by using a common scalar process (the regional wind condition) for the different locations. This scalar process, denoted $\{X_t\}$ in the sequel, is not directly observable and is introduced as a latent process.  Due to the mean motion of air masses we expect that the wind conditions at western locations will depend more on the leading one lag $X_{t+1}$ and $X_{t}$ signals than on the lagged signal $X_{t-1}$ with the reverse phenomenon at eastern locations. These points led us to consider
 the following Gaussian state-space model
$$
(\M)\left \{
 \begin{array}{lcl}
X_{t+1} &=& \rho X_{t} + \sigma \epsilon_{t+1},\\
Y_{t} &=& \alpha_{1} X_{t+1} + \alpha_{0} X_{t} + \alpha_{-1} X_{t-1} + \Gamma^{1/2} \eta_{t}
 \end{array}
\right.\quad \textrm{for $t\geq 0$.}$$
$Y_t \in \mathbb{R}^{K}$ is the observed process, its $K$ coordinates correspond to the mean corrected transformed wind speed at the $K=18$ locations.  $\{\epsilon_t\}$ and $\{\eta_t\}$ are independent Gaussian white noise sequences with zero means and identity covariance matrices. $\alpha_{1}$, $\alpha_{0}$ and $\alpha_{-1}$ are $K$-dimensional vectors which link the lagged values of the regional process $\{X_t\}$ to local wind conditions. The covariance matrix $\Gamma \in \mathbb{R}^{K \times K}$ models the spatial structure of the difference between the observed process $\{Y_t\}$ and the local conditions $W_{t} = \alpha_{1} X_{t+1} + \alpha_{0} X_{t} + \alpha_{-1} X_{t-1}$ deduced from the regional process. It  may correspond to small scale fluctuations.  In finance and economics this covariance matrix of error of measurement is often diagonal. In \cite{WIK10} a parametrization of this matrix or the use of a diagonal matrix are advised. As a first step this covariance matrix is not parameterized but reduced models which take into account the spatial information are investigated in Subsection \ref{paramGamma}. In the sequel we denote  $\Lambda = (\alpha_{1} | \alpha_{0} | \alpha_{-1}) \in \mathbb R^{K \times 3}$  and $\theta = \left(\rho,\sigma,\Lambda,\Gamma \right)$ the unknown parameter.

The temporal dynamics of the observed process is mainly contained in the latent process $\{X_t\}$. The model thus imposes the same long term temporal dynamics, corresponding to the regional scale, at each location. Under the assumption $\abs{ \rho } < 1$, the AR(1) process $\{X_t\}$ is stationary and so is the process $\{Y_t\}$.  $\{W_{t}\}$ is an ARMA(1,2) process since
$$W_{t} - \rho W_{t-1}= \alpha_{1} \epsilon_{t+1} + \alpha_{0} \epsilon_{t} + \alpha_{-1} \epsilon_{t-1}.$$
Signs and values of $(\alpha_{1},\alpha_{0},\alpha_{-1})$ can be interpreted in terms of autocovariance function of the moving average part $\alpha_{1} \epsilon_{t+1} + \alpha_{0} \epsilon_{t} + \alpha_{-1} \epsilon_{t-1}$.



\subsection{Second order structure and identifiability}\label{cov-identif}

Identifiability is required to get sensible and reliable parameter estimation. Gaussian linear state-space models are often non-identifiable without additional constraints  (see e.g. \cite{HANN88}, \cite{LJUN99}, \cite{BAI12}, \cite{BOR10}). Indeed the introduction of the latent process $\{X_t\}$ is a source of non-identifiability since the unknown parameters need to be identified uniquely from the distribution of the observed $\{Y_t\}$. 

 $\{Y_t\}$ is a zero mean stationary Gaussian process which is thus characterized by its second order structure given below
\begin{eqnarray}
\nonumber \cov_{\theta}(Y_{t},Y_{t}) & = & \frac{\sigma^2}{1-\rho^2}\Big( \alpha_{1} (\alpha_{1} + \rho \alpha_{0} + \rho^{2} \alpha_{-1})^{t} + \alpha_{0} (\rho \alpha_{1} + \alpha_{0} + \rho \alpha_{-1})^{t} +\\
& &  \alpha_{-1} (\rho^{2}\alpha_{1} + \rho \alpha_{0} + \alpha_{-1})^{t}\Big) + \Gamma,\label{cov0}\\
\nonumber \cov_{\theta}(Y_{t},Y_{t+1}) & = & \frac{\sigma^2}{1-\rho^2}\Big(\alpha_{1} (\rho \alpha_{1} + \alpha_{0} + \rho \alpha_{-1})^{t} + \alpha_{0} (\rho^{2}\alpha_{1} + \rho \alpha_{0} + \alpha_{-1})^{t} +\\
& & \rho \alpha_{-1} (\rho^{2}\alpha_{1} + \rho \alpha_{0} + \alpha_{-1})^{t}\Big),\label{cov1}\\
\cov_{\theta}(Y_{t},Y_{t+k}) & = & \frac{\sigma^2}{1-\rho^2}\rho^{k-2}(\alpha_{1} + \rho \alpha_{0} + \rho^{2} \alpha_{-1})(\rho^{2}\alpha_{1} + \rho \alpha_{0} + \alpha_{-1})^{t},\label{cov2}\\
\nonumber & & \textrm{\hspace{.1cm}for all $k\geq 2$}. 
\end{eqnarray}

The study of this space-time covariance function leads to the following Proposition which is proven in Appendix \ref{nonidentif}.

\begin{prop} 
\label{propid}
Assume that $(\M)$ holds. Assume further that  $\frac{\sigma^2}{1-\rho^2}=1$ and that the vectors $\alpha_{1}$, $\alpha_{0}$ and $\alpha_{-1}$ are linearly independent. Then the parameters can be identified from the distribution of the process $\{Y_t\}$.
\end{prop}

These identifiability constraints were always satisfied when fitting the model to the data. The first condition requires $X$ to have a unit stationary variance, the local variability is then accounted in the vectors $\alpha_{1}$, $\alpha_{0}$ and $\alpha_{-1}$. The second one is natural since in case of linear dependence between these vectors the model (M) is reduced to a sub-model dealing with one or two lagged versions of $X$. Identifiability of linear Gaussian state-space models have been investigated during the last decades and initially in control theory. Literature is abundant on stochastic linear systems identification (\cite{LJUN99,HANN88}). To the best of our knowledge most of sufficient conditions of identifiability are structural constraints on parameters (see \cite{PAPA10} for references and examples) associated with identification procedures.  Identifiability is examined through different criteria such that transfer functions (\cite{LJUN99}) or likelihood criterion like in \cite{PAPA10}. Identification procedures are realized through controllability and observability of several parameters in \cite{LJUN99} and via the EM algorithm in \cite{PAPA10}. In econometrics the identifiability of the latent factors and the loading matrix is considered (see for example \cite{BAI12,BOR10}). In most cases structural constraints are also applied depending on the interpretability desired. However the general conditions given in \cite{BAI12} do not guarantee identifiability of the model (M) since $X$ is scalar in (M).

Considering the function defined by  (\ref{cov0}-\ref{cov2})  as a discretized space-time covariance function, we show in Appendix \ref{nonidentif} that properties of full-symmetry and separability are not fulfilled under the identifiability constraints of Proposition \ref{propid}. Other non-symmetric space-time covariance models have been proposed in the literature and some of them have been adjusted to the Irish wind dataset that exhibits non-symmetry (see \cite{GNEI02} and \cite[chapter 4]{GNEI07}). Strong spatial assumptions are commonly assumed in these models such as spatial stationarity and isotropy. However due to prevailing flows a lot of meteorological data have anisotropic patterns like the dataset under study or the Irish one. A model, based on the specification of a vector autoregressive process, is proposed in \cite{LUN05} to capture a part of the anisotropy that is observable in the correlations of the Irish dataset. We will see in Section \ref{results} that the proposed model enables to reproduce various above mentioned complex properties of the space-time covariance of our wind data.

\subsection{Parameter estimation}\label{estim}

Two methods of estimation have been implemented and compared. The first one is a method of moment based on the second order structure of the process $\{Y_t\}$ given by (\ref{cov0}-\ref{cov2}). It consists in minimizing numerically the following objective function
\begin{eqnarray}
\label{eq:obj} \theta & \rightarrow & \| \widehat{\cov}(Y_{t},Y_{t}) - \cov_{\theta}(Y_{t},Y_{t})\|_{2}^{2} +  \|\widehat{\cov} (Y_{t},Y_{t+1}) - \cov_{\theta}(Y_{t},Y_{t+1})\|_{2}^{2} \\
\nonumber & & + \|\widehat{\cov}(Y_{t},Y_{t+2}) - \cov_{\theta}(Y_{t},Y_{t+2})\|_{2}^{2} + \|\widehat{\cov} (Y_{t},Y_{t+3}) - \cov_{\theta}(Y_{t},Y_{t+3})\|_{2}^{2},
\end{eqnarray}

where $\widehat{\cov}$  denotes the empirical covariance function and $\|.\|_{2}$ stands for the matrix Frobenius norm. This method, denoted GMM in the sequel, is a standard method in geostatistics (see e.g. \cite{CRES91}). We have chosen to consider only the first four lags of the autocovariance function when building the objective function (\ref{eq:obj}). It corresponds to the minimal number of terms needed to identify the parameters (see Appendix \ref{nonidentif}). Simulation results indicate that including more lags in the objective function does not lead to more accurate estimates. 

The second method performs Maximum Likelihood (ML) estimation  using the Expectation-Maximization (EM) algorithm (\cite{CAP05} and \cite{DEM77}). For linear Gaussian state-space model efficient numerical procedures exist for both the E-step, where the Kalman recursions lead to an exact computation of the smoothing probabilities, and the M-step with analytical expressions being available for the parameters which maximize the intermediate function of the EM algorithm. More details can be found in the supplementary materials.

Both methods are sensitive to the  initial parameter value which needs to be chosen carefully. We used the following procedure which exploits the properties of the second order structure of $\{Y_t\}$:
\begin{itemize}
\item[-] $\rho= \displaystyle \frac{\cov(Y_{t},Y_{t+3})_{i,j}}{\cov(Y_{t},Y_{t+2})_{i,j}}$ for all $i,j \in \{1,...,K\}$ is initialized as the empirical mean of  $\displaystyle \frac{\widehat{\cov}(Y_{t},Y_{t+3})_{i,j}}{\widehat{\cov}(Y_{t},Y_{t+2})_{i,j}}$.
\item[-]  $\Lambda$ is estimated by minimizing
\begin{eqnarray}
\nonumber
\theta_{\Lambda} & \rightarrow &\|\widehat{\cov}(Y_{t},Y_{t+1}) - \cov_{\theta}(Y_{t},Y_{t+1})\|_{2}^{2}+ \|\widehat{\cov}(Y_{t},Y_{t+2}) - \cov_{\theta}(Y_{t},Y_{t+2})\|_{2}^{2} 
\end{eqnarray}
as a function of $\Lambda$ with $\rho$ being fixed to the value obtained in the previous step. Note that this function does not depend on $\Gamma$ according to \eqref{cov1} and \eqref{cov2}. 
\item[-] $\Gamma$ is determined by  minimizing
\begin{eqnarray}
\nonumber
\theta_{\Gamma} & \rightarrow &\|\widehat{\cov}(Y_{t},Y_{t}) - \cov_{\theta}(Y_{t},Y_{t})\|_{2}^{2}
\end{eqnarray}
as a function of $\Gamma $ with $\rho$ and $\Lambda$ being fixed to the value obtained in the previous steps.
\end{itemize} 
These rough estimates are used as initial conditions for the numerical optimization of \ref{eq:obj} which leads to the GMM estimates. The GMM estimates are then used to initialize the EM algorithm and get the ML estimates.

%
%


\subsection{Properties of the estimates}

Under suitable  conditions, GMM (see \cite{NEW94}) and ML (see \cite{NEW94, SHU06,HANN88,CAI88}) estimators are consistent and asymptotically Gaussian. A simulation study was performed in order to assess the performance of the estimates in a situation comparable to the practical application. $N=100$ independent sets of the size of the studied data are simulated for the parameters set estimated by ML on the wind data. Table \ref{prop_param} gives the bias, standard deviation and Root Mean Square Error (RMSE) of ML and GMM estimates computed from the simulations. Bias and standard deviations are low
. ML generally outperforms GMM except for estimating $\Gamma$ where both methods give comparable results. For both methods $\alpha_{1}$ and $\alpha_{-1}$ are more accurately estimated than $\alpha_{0}$ and $\Gamma$ is the less accurately estimated quantity.


\begin{table}
\begin{center}
\scalebox{.7}{
\noindent\begin{tabular}{|c|c|c|c|c|c|c|}
\hline
\multicolumn{1}{|c|}{}  & \multicolumn{2}{|c|}{ Bias } & \multicolumn{2}{|c|}{ Sd} & \multicolumn{2}{|c|}{ RMSE}\\
\hline
\multicolumn{1}{|c|}{Parameters}  & \multicolumn{1}{|c|}{ GMM } & \multicolumn{1}{|c|}{ML} & \multicolumn{1}{|c|}{ GMM } & \multicolumn{1}{|c|}{ML} & \multicolumn{1}{|c|}{ GMM } & \multicolumn{1}{|c|}{ML}\\
\hline
{$\rho$}  &  0.036 & 0.004 & 0.022 &  0.017 &  0.042 &  0.017\\
\hline
$\alpha_{1}$ & [-0.11;-0.009] & [-0.069;-0.019] & [0.065;0.108] & [0.071;0.097] & [0.067;0.149] & [0.068;0.127] \\
\hline
$\alpha_{0}$ & [-0.047;-0.234] & [0.054;0.144] & [0.11;0.182] & [0.11;0.144] & [0.125;0.292] & [0.127;0.228]\\
\hline
$\alpha_{-1}$ & [-0.080;0.022]  & [-0.035;0.012] & [0.078;0.114] &  [0.062;0.104] & [0.086;0.139] & [0.079;0.117]\\
\hline
$\Gamma$ & [-0.199;0.007] & [-0.108;0.013] & [0.058;0.367] & [0.029;0.368]  & [0.053;0.199] & [0.053;0.115] \\
\hline
\end{tabular}}
\caption{Bias, standard deviation and RMSE of parameters estimates. For the multidimensional parameters, minimal and maximal values are given in brackets. }
\label{prop_param}
\end{center}
\end{table}


\section{Results}\label{results}

In order to validate the proposed model we checked its physical realism, its ability to generate realistic wind conditions and to produce accurate forecasts. GMM and ML estimates are also compared through this validation in order to check their robustness in a practical context.

\subsection{Interpretability}
\label{sec:inter}

\begin{figure}
\begin{center}
\includegraphics[width=4.5cm]{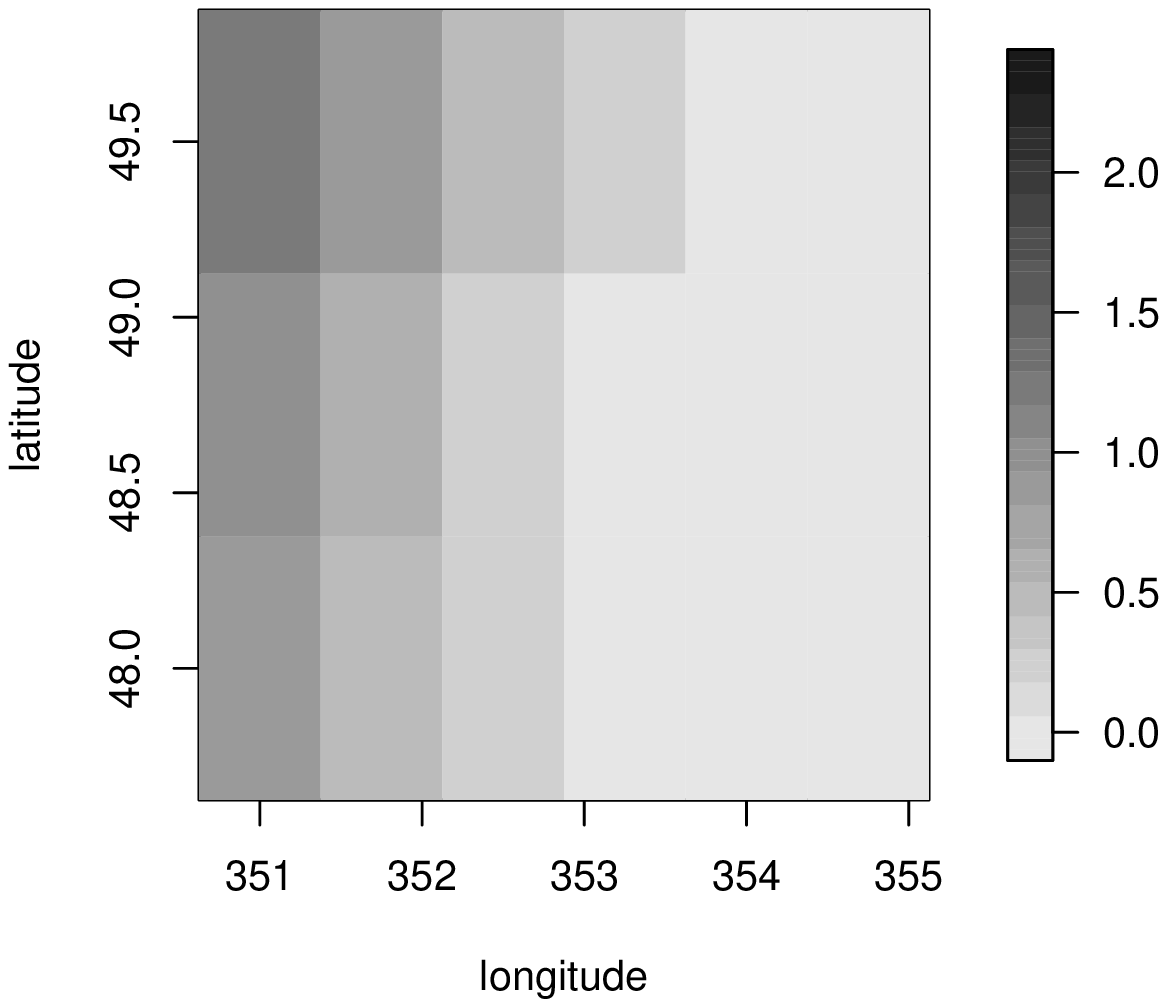}
\hglue.0001cm
\includegraphics[width=4.5cm]{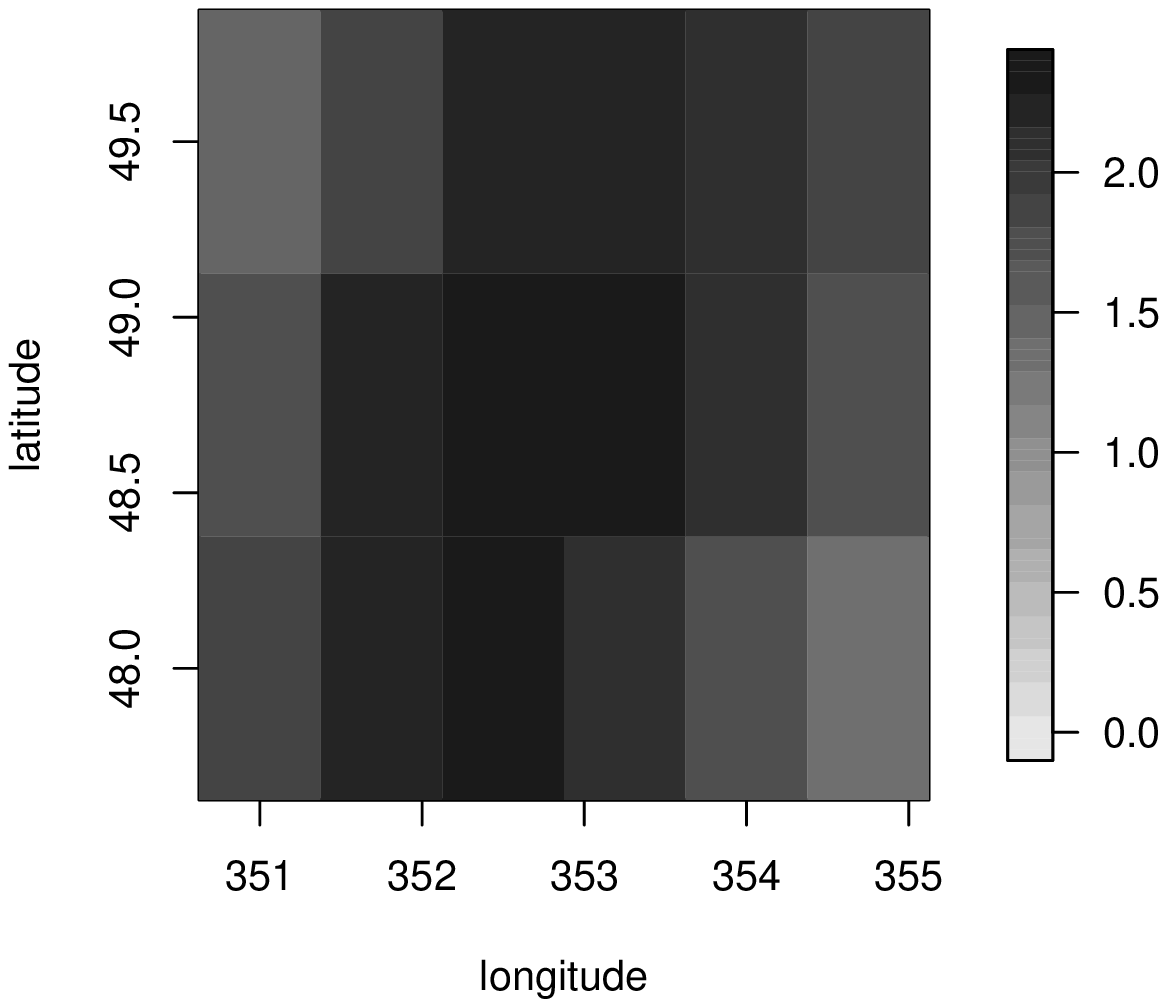}
\hglue.0001cm
\includegraphics[width=4.5cm]{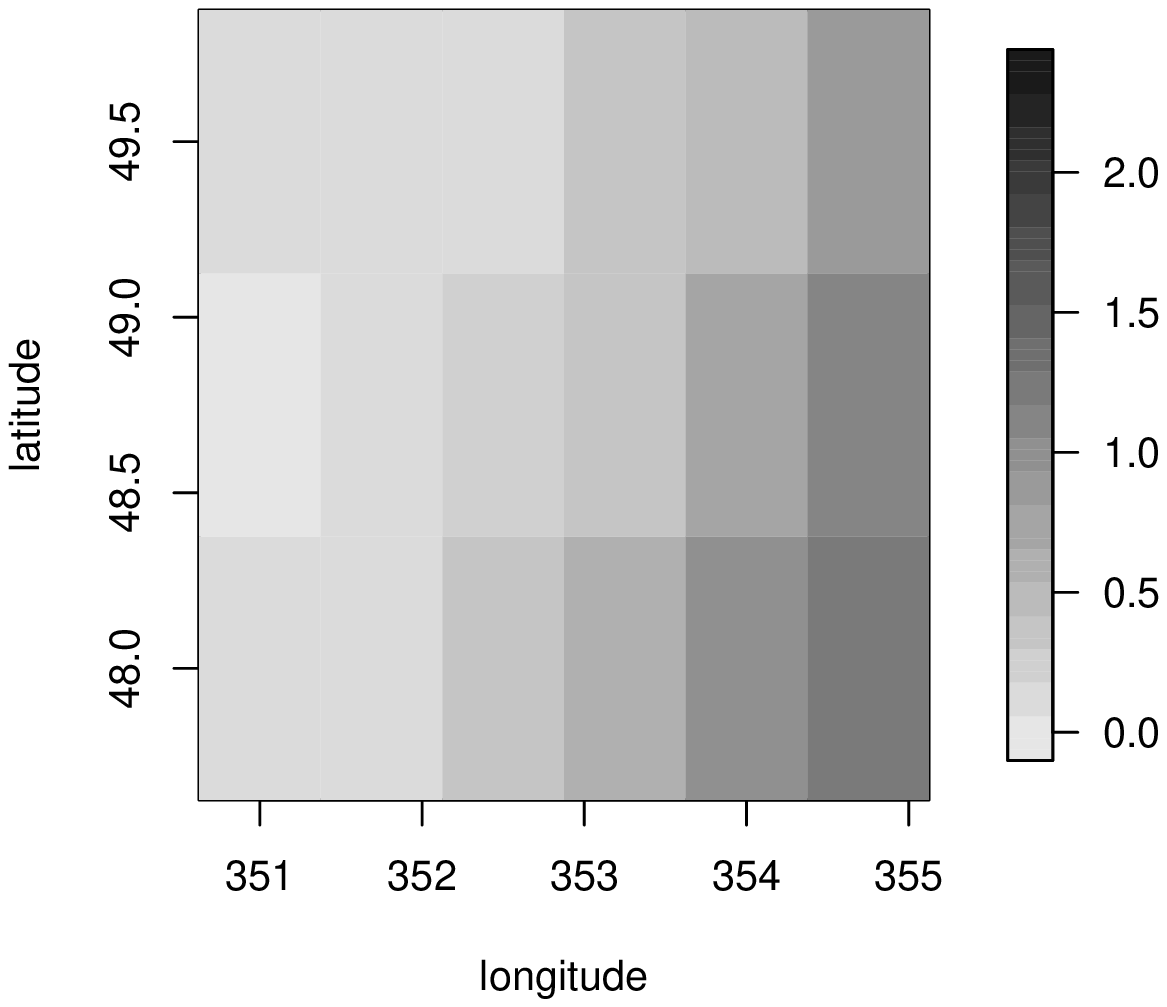}
\caption{ML estimate of $\alpha_{1}$ (left panel) $\alpha_{0}$ (middle panel) and $\alpha_{-1}$ (right panel).}
\label{imageLambda}
\end{center}
\end{figure}

The loading matrix $\Lambda$ links the latent process to observed wind conditions. The values of $\alpha_{1}$ and $\alpha_{-1}$ shown on Figure \ref{imageLambda} reveal the site-dependent relations with the regional mean wind. As expected western locations, which are generally the first affected when meteorological events enter in the studied region, depend more on $X_{t+1}$ than on $X_{t-1}$ and the reverse is true for eastern locations.

\begin{figure}
\begin{center}
\includegraphics[width=6.5cm]{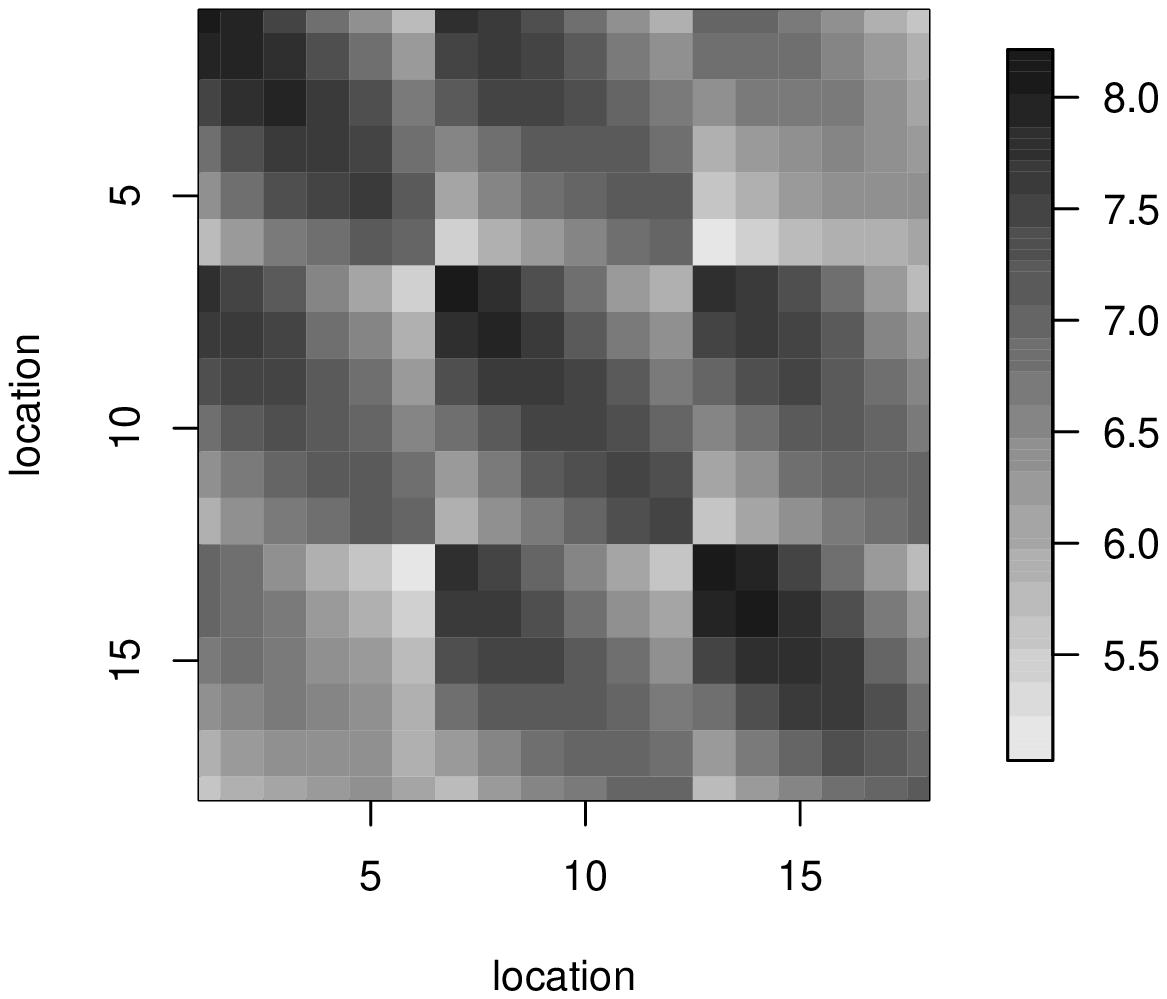}
\hglue.001cm
\includegraphics[width=6.5cm]{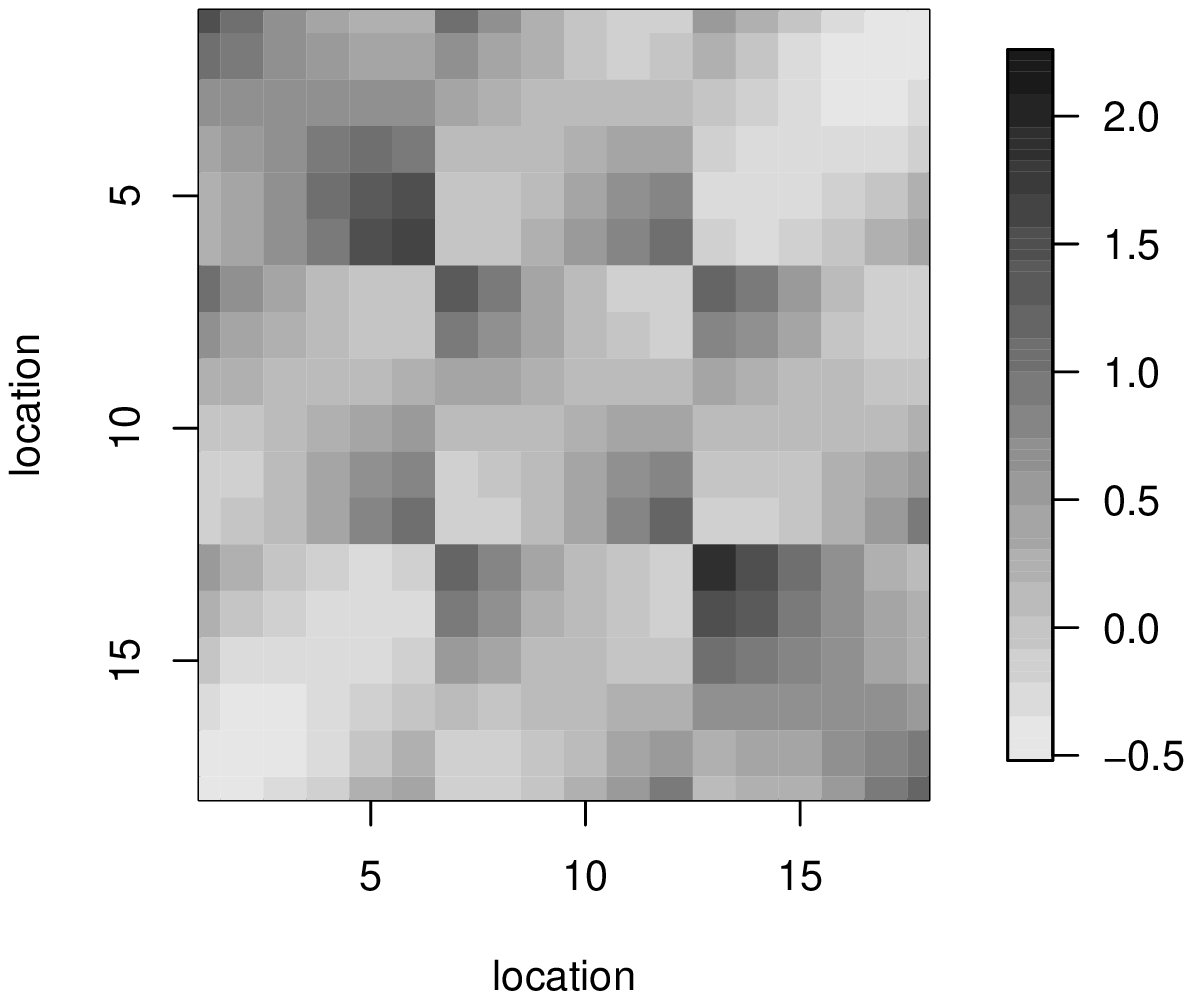}
\caption{Covariance matrix of $Y$ (left) and ML estimate of $\Gamma$ (right).}
\label{imageGamma}
\end{center}
\end{figure}

Since large scale variability is supposed to be contained in the latent process,  $\Gamma$ should contain only small scale variations due to the differences between the observed wind $\{Y_t\}$ and the downscaled regional wind $\{W_t\}$. This is confirmed when comparing the spatial sill and range of $\Gamma$ with the ones of the original covariance function of the  data (see Figure \ref{imageGamma}). 
The shape of $\Gamma$ has a block structure  which is induced by the geometry of the domain and the numbering of the sites (see Figure \ref{map}). The level sets of the blocks, except the top right corner (and by symmetry bottom left corner), are similar to saddle point level sets: the model better explains the wind observed at the central locations of the domain than at the locations which are close to the boundary. The top right corner has elliptical level sets. This different geometry raises problems when trying to develop simple parametric model for $\Gamma$ (see Section \ref{paramGamma}).

\subsection{Realism of simulated sequences}
\begin{figure}
\begin{center}
\includegraphics[width=6.5cm]{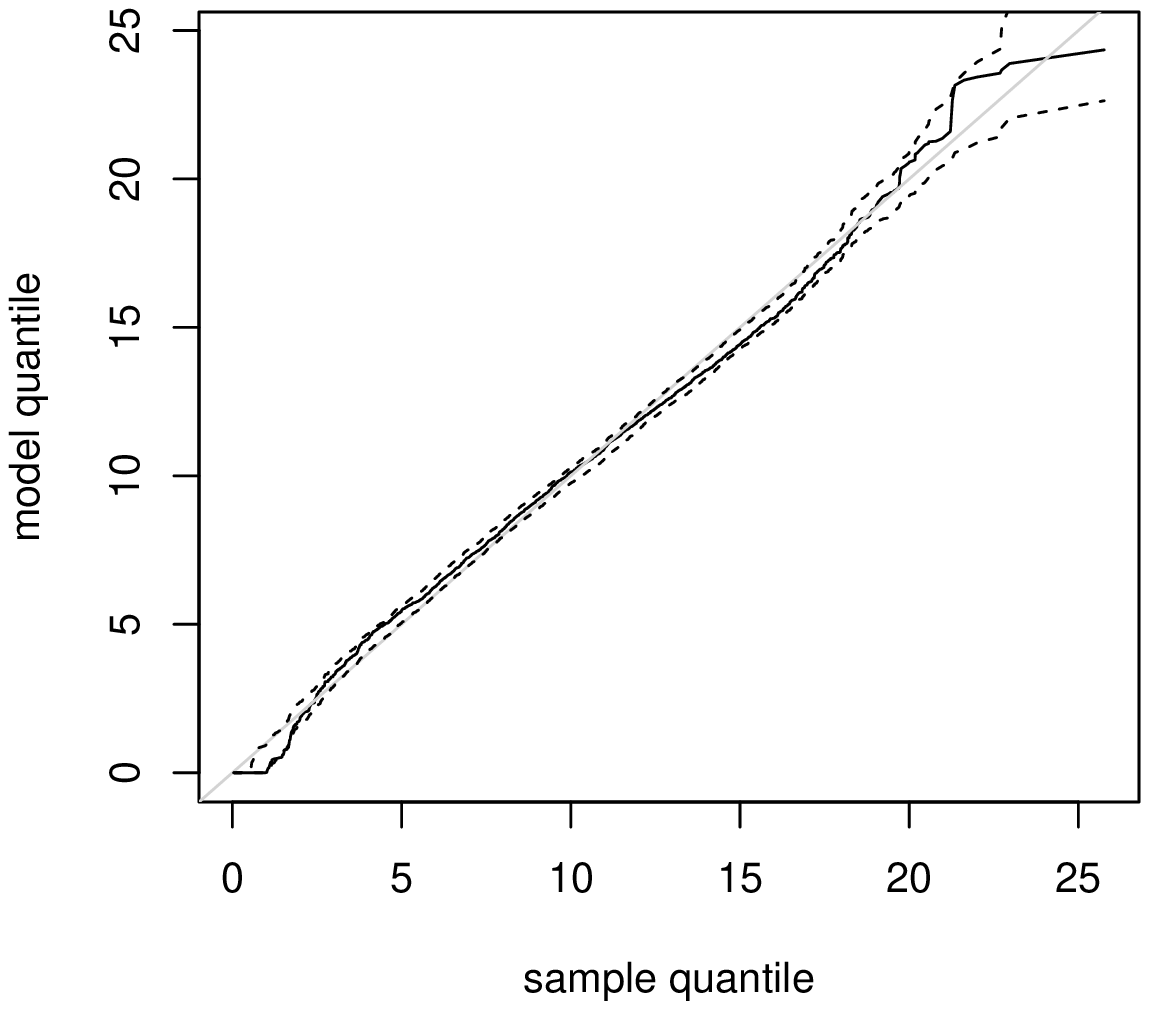}
\hglue .1cm 
\includegraphics[width=6.5cm]{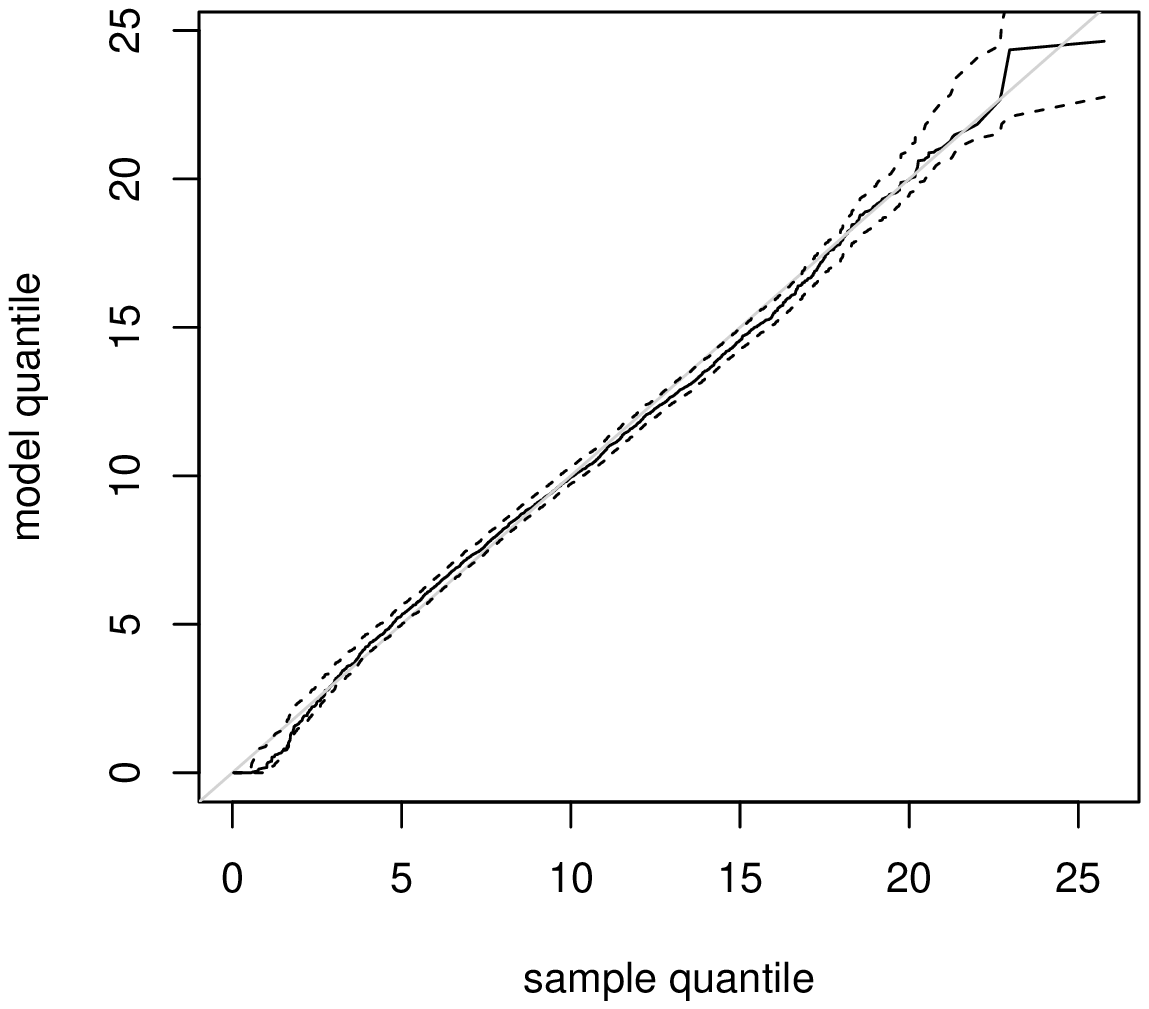}
\caption{Quantile-Quantile plot at location $9$ for the model ($\M$) and the parameters estimated by GMM (left) and by ML (right). The dashed lines corresponds to  90\%
prediction intervals computed by simulation.}
\label{marge}
\end{center}
\end{figure}

To further validate the model we have checked its ability to simulate realistic wind conditions. For that we have generated artificial time series from the model, and have compared
statistics corresponding to the artificial sequences with those from the original data. We first looked at the marginal distributions of wind speed, both observed and fitted, at each
location. According to quantile-quantile plots shown on Figure \ref{marge}, the model is able to reproduce the general shape of the marginal distribution of the process at the central station $9$ except for very low wind speed. Similar results were obtained at other locations. 

\begin{figure}
\begin{center}
\includegraphics[width=6.5cm,angle=90]{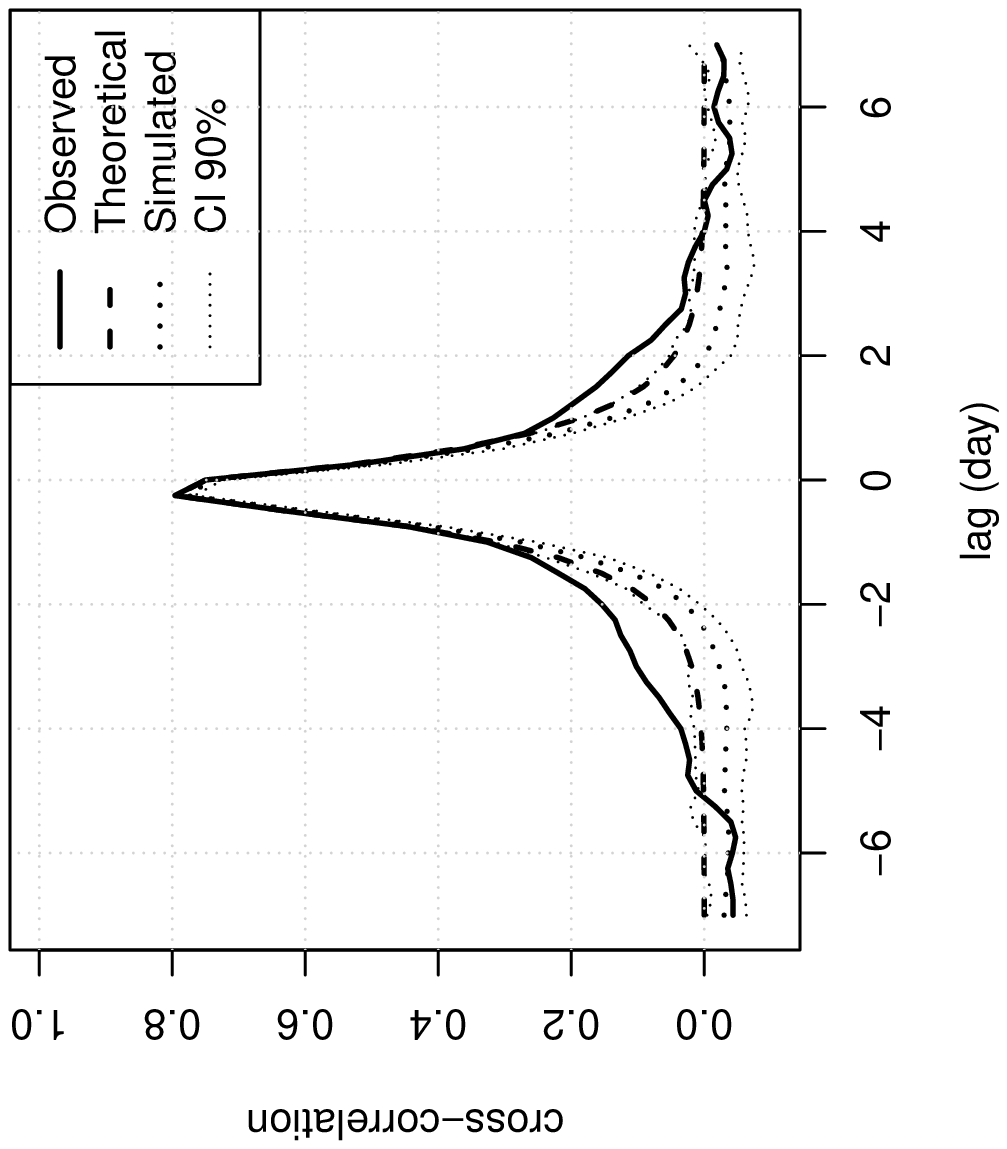}
\hglue .1cm 
\includegraphics[width=6.5cm,angle=90]{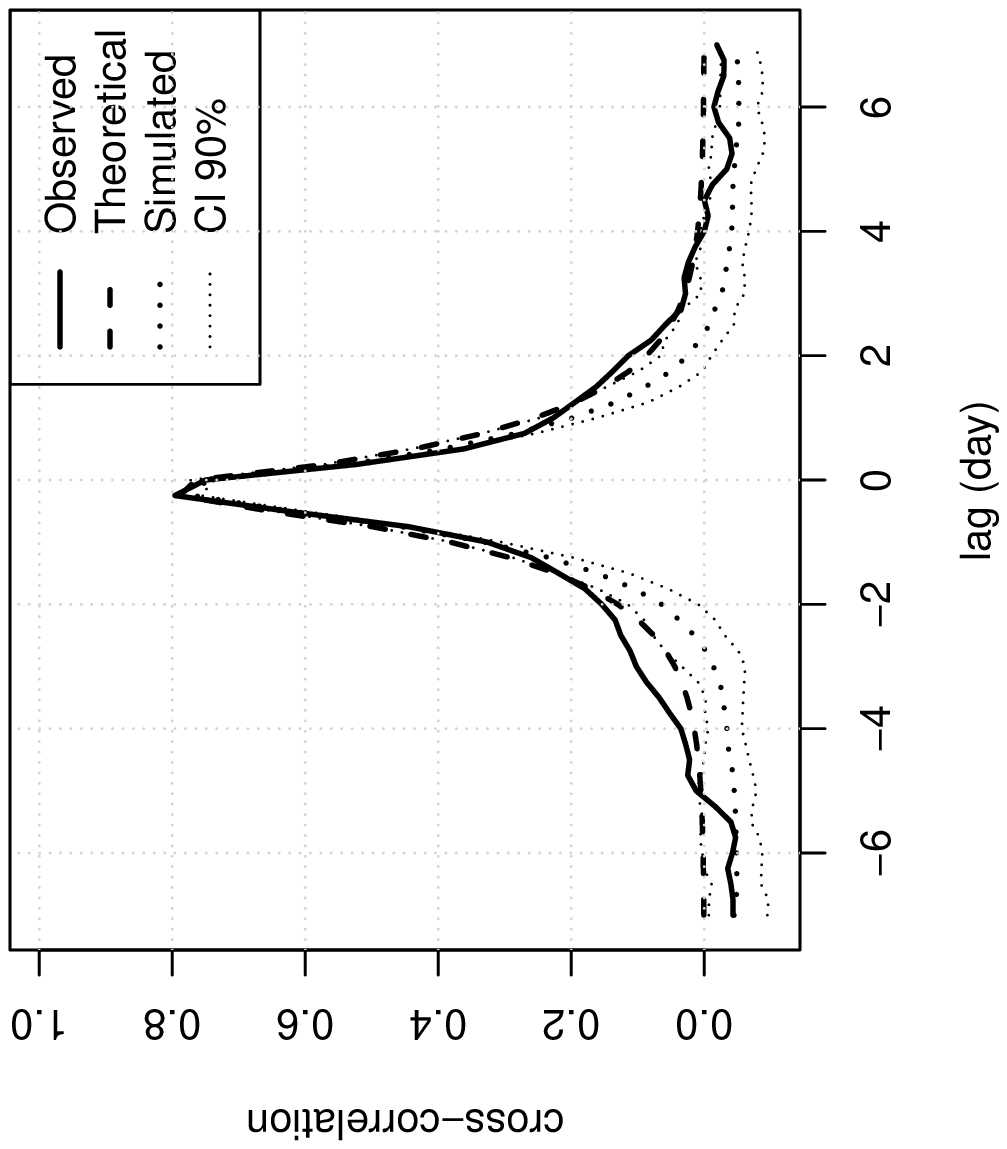}
\vglue.01cm
\includegraphics[width=6.5cm]{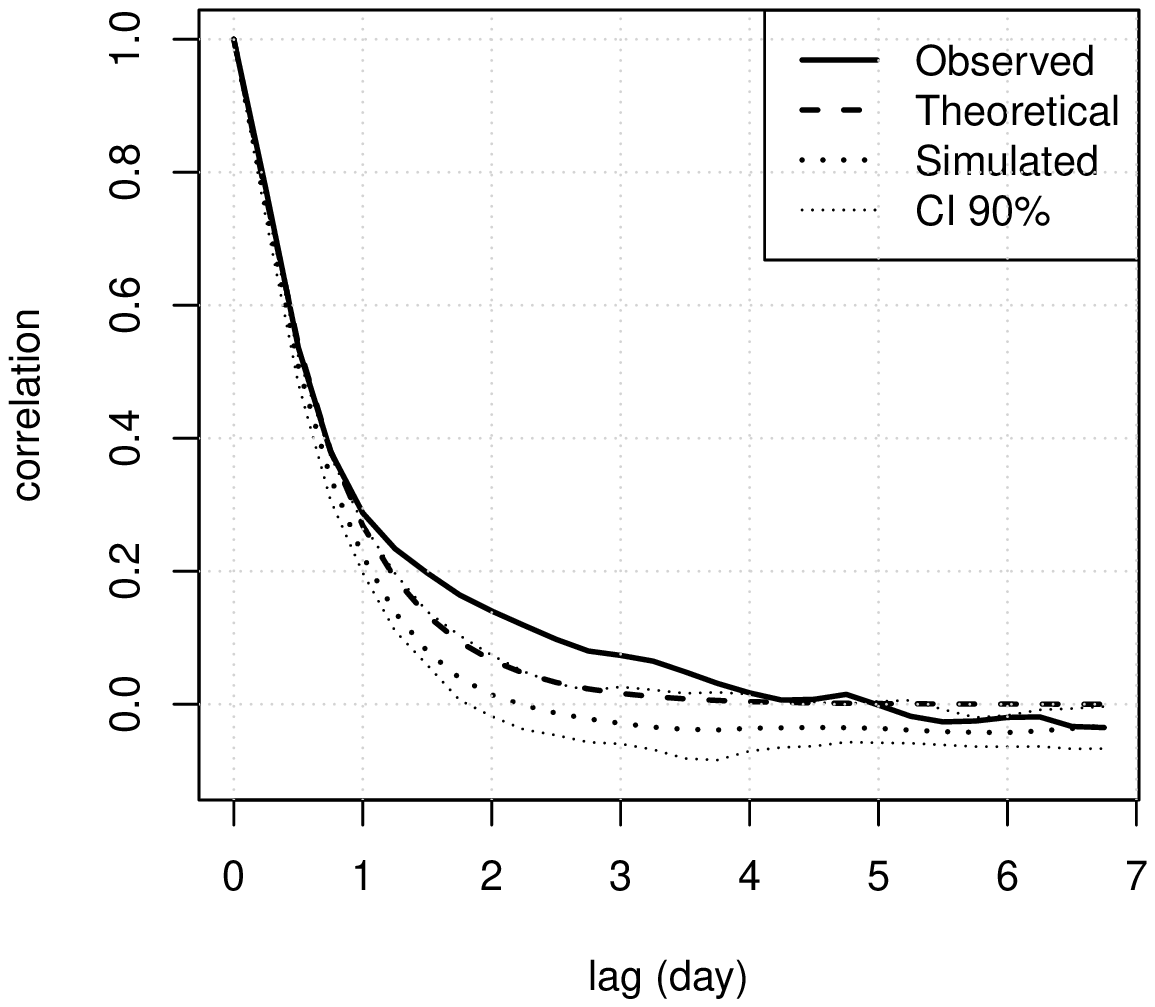}
\hglue .1cm 
\includegraphics[width=6.5cm]{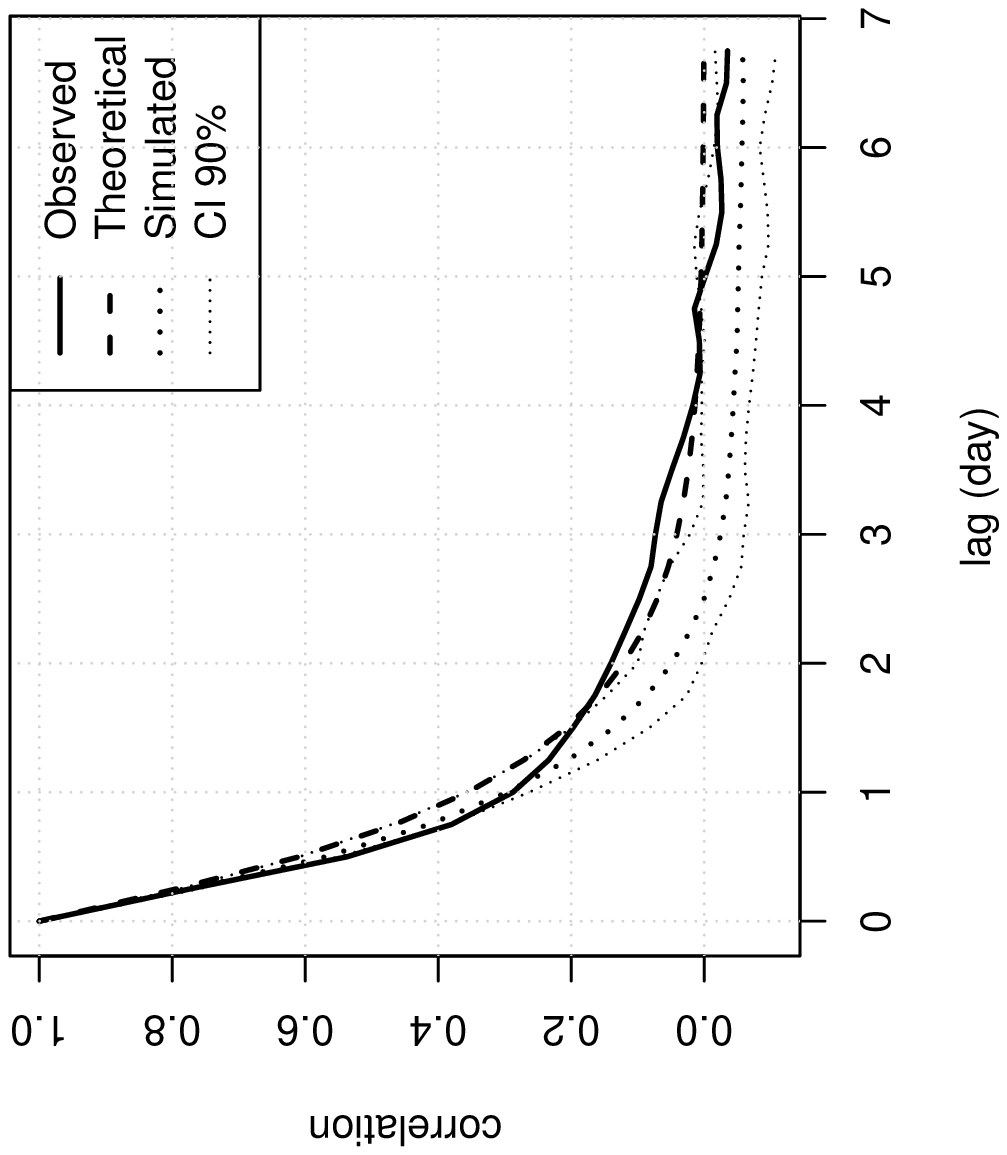}
\caption{Observed (full lines) and theoretical (dashed lines) cross-correlations between locations 13 and 18 (upper line) and auto-correlation at location $9$ (lower line) for the model ($\M$) with parameters estimated by GMM (left) and by ML (right). 90\% prediction intervals are computed from $100$ independant samples of the size of the original data.
}
\label{crosscorr1318}
\end{center}
\end{figure}

The model assumes that the dynamics of wind speed are inherited mainly from the common latent process $\{X_t\}$. To check this assumption, we have compared the autocorrelation functions of the model to the data ones at the various locations. Typical results are shown in Figure \ref{crosscorr1318}. The model is able to reproduce the first coefficients of the autocorrelation function at the central location $9$ and of the cross-correlation function between locations $13$ and $18$ which exhibits a time shift due to the prevailing westerly flow. GMM estimate is designed to make coincide the first lags of the empirical autocovariance functions with the one of the fitted model. Figure \ref{crosscorr1318} shows that the agreement is indeed very good. However the ML method, which takes into account the longer term dynamics in the likelihood function, is better in reproducing the correlation structure for time lags above one day. It is mainly due to the higher value of $\rho$ ($0.76$ by ML and $0.70$ obtained by GMM). The better performance of ML estimates is coherent with  Table \ref{prop_param}. Figure \ref{crosscorr1318} also shows the important bias on the estimates of the second order structure of the process which is problematic when using the GMM approach.

\begin{figure}
\begin{center}
\includegraphics[width=6.5cm]{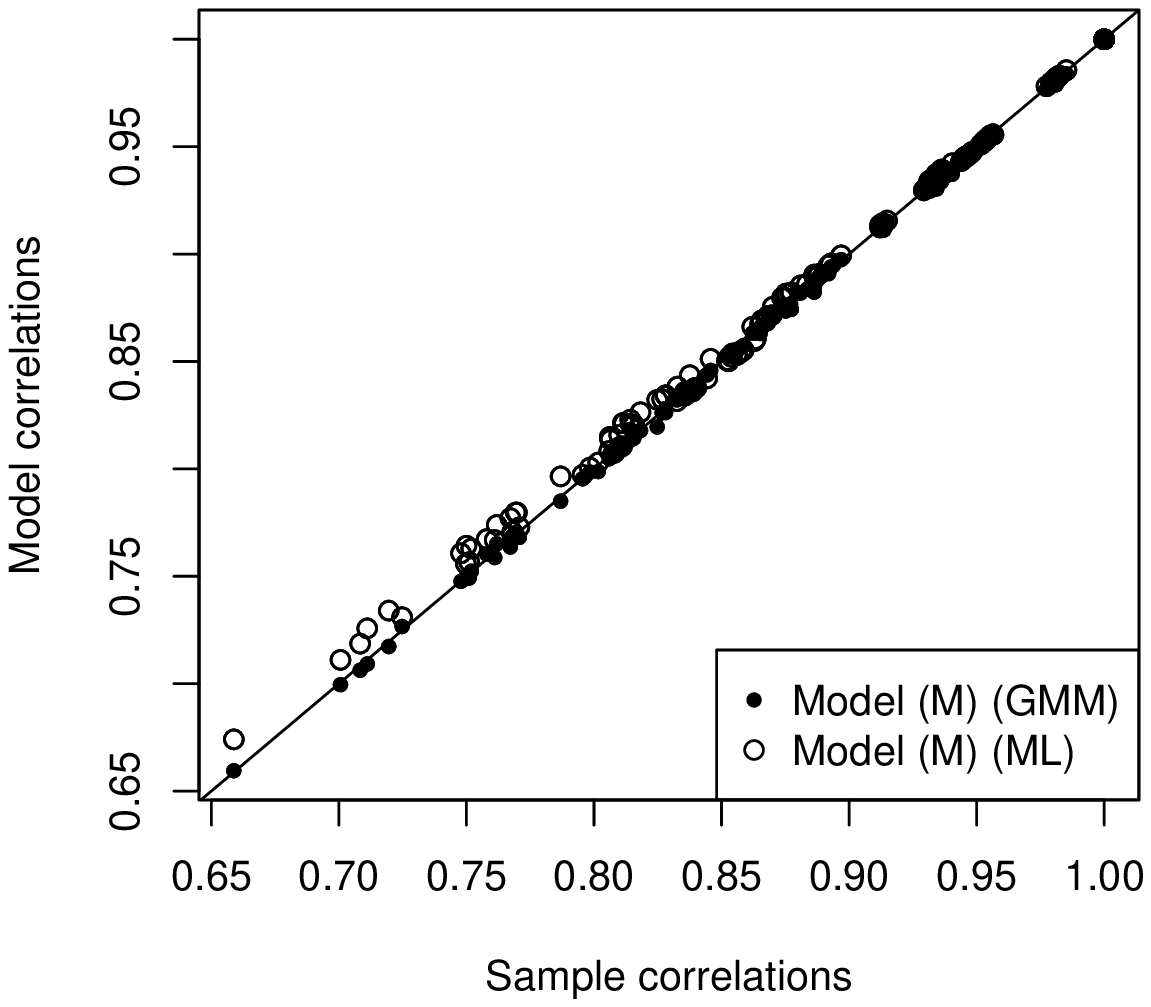} 
\hglue.0001cm
\includegraphics[width=6.5cm]{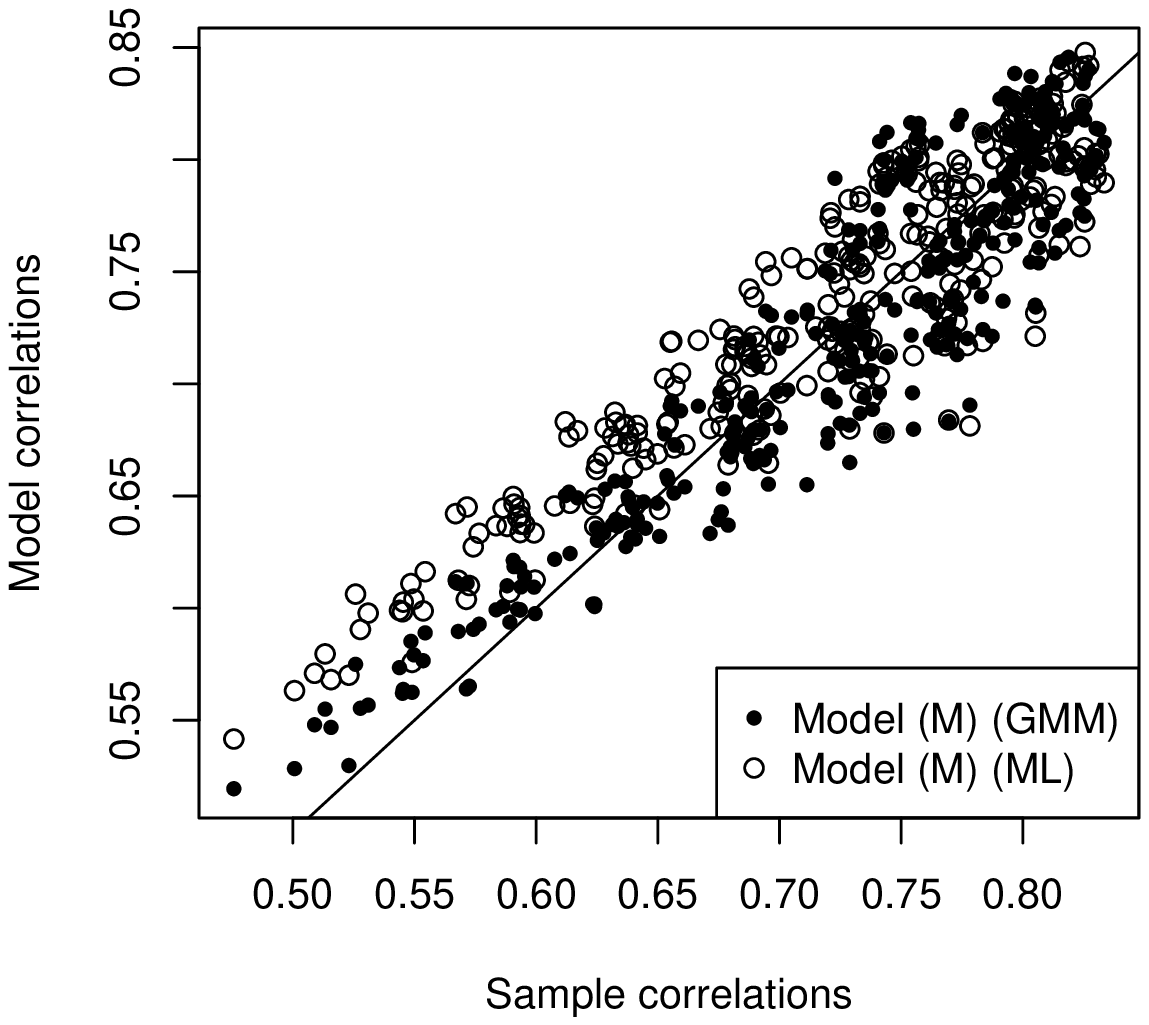} 
\caption{Theoretical correlations against observed correlations at lag $0$ (left) and lag $1$ (right) for the model ($\M$) and the two methods of estimation.}
\label{corr0&1}
\end{center}
\end{figure}

Figure \ref{corr0&1} shows the theoretical correlations at lags $0$ and $1$ against the empirical ones to assess the model's ability to capture the spatial dependence structure. Correlations at lag 0 are very well reproduced and lagged one correlations are also generally well reproduced. As expected, the GMM method provides again a better fit. Similar figures than Figure \ref{corrVSlatlong} reveal that the anisotropy and non-separability observable in the correlations are partly reproduced by the model.

\subsection{Prediction}

An advantage brought by the Markovian structure of the model ($\M$) is that short term forecast can be efficiently computed through the Kalman recursions (see \cite[chapter 8]{BROC06}). The forecast performance is assessed by computing the one-step ahead forecast on the last $8$ years of data (validation set) after fitting the model on the first $25$ years of data (training set). In practice the forecast skill of the model at location $i \in \{1,...,K\}$ is evaluated by computing the natural empirical estimate of the Mean Square Percentage Error (MSPE) defined as
$$\textrm{MSPE}(i) = \frac{\var (Y_t(i)-\E[Y_t(i)|Y_0,...,Y_{t-1}])}{\var (Y_t(i))}$$
where the MSE of the forecast error (the numerator) is normalized by the variance of the field at the individual locations, with $Y_t$ the original non transformed wind.  

According to Table \ref{BIC}, model ($\M$) leads to significant improvements over persistence forecast (about 28\%), which is the classical benchmark for wind prediction (see e.g. \cite{ZHU12}), and the forecast obtained when fitting a different ARMA(2,1) model at each site (about 17\%). It illustrates the gain of using the spatial information when computing short-term probabilistic forecasts. For comparison purpose, a vector autoregressive model of order $1$ (VAR(1)) was also fitted. The VAR(1) model gives slightly better results compared to model ($\M$) with an improvement of about $9\%$ in average over all locations. However the VAR(1) model involves $495$ parameters, compared to the $208$ parameters for model ($\M$), and the VAR(1) parameters are difficult to interpret. 

\begin{figure}
\begin{center}
\includegraphics[width=7cm,angle=90]{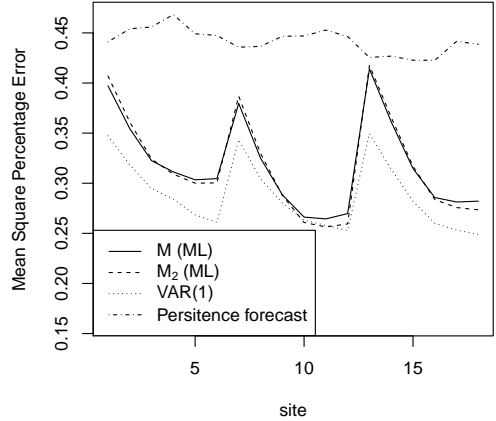}
\caption{ Mean Square Percentage Error (MSPE) between observed wind speed and back transformed predicted wind speed at each location for models ($\M$), ($\M_{2}$), VAR(1) and persistence.}
\label{rmse}
\end{center}
\end{figure}

The spatial structure of the forecast error is shown on Figure \ref{rmse}. The improvement obtained with the VAR(1) and ($\M$) models, which include the spatial information, is more pronounced at eastern locations. It is due to the prevailing westerly flow. Indeed the wind speed observed at the western locations ($1,2,7,8,13,14$) at given time $t$ brings information on the wind speed observed at the eastern locations ($5,6,11,12,17,18$) at the next time $t+1$. Note also that the difference between models ($\M$) and VAR(1) is very low at the central locations $9, 10,11$ but the forecast performance of model ($\M$) tends to deteriorate close to the boundaries of the studied region. 
We observed that the forecasts of model ($\M$) exhibit less spatial variability than the observations and the forecasts obtained with the VAR(1) model. It suggests that using of a scalar latent process $X$ is too simplistic to catch all the complexity of the space-time structure of the data and that $X$ is designed mainly to describe the wind conditions at central locations.

\section{Some improvements of the model}\label{parsimonious model}

In this section we explore reduced models for the matrices $\Gamma$ and $\Lambda$ and higher order models for $\{X_t\}$.


\subsection{Parameterization of $\Gamma$}\label{paramGamma}

\begin{figure}
\begin{center}
\includegraphics[width=6.5cm]{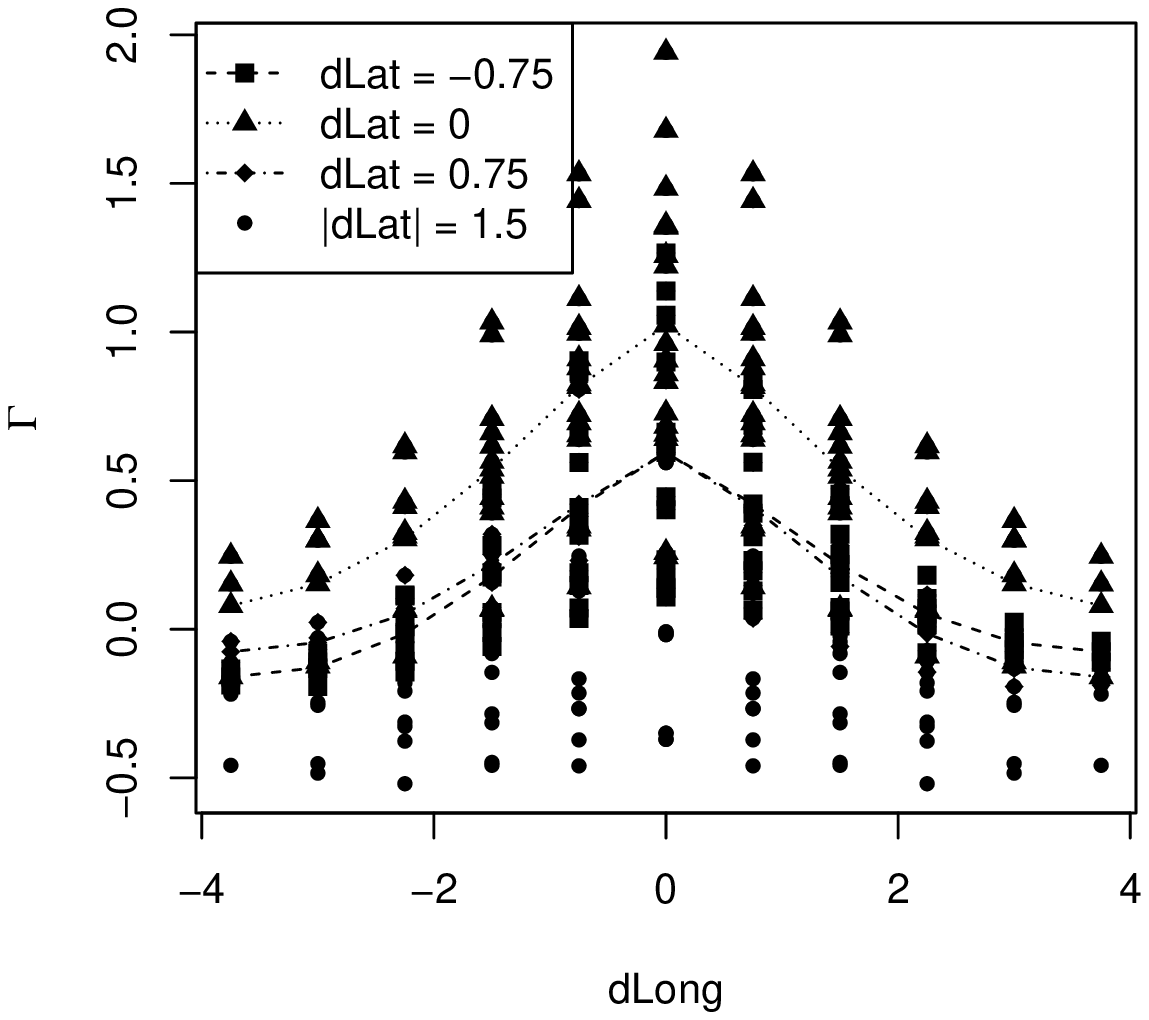}
\hglue.001cm
\includegraphics[width=6.5cm]{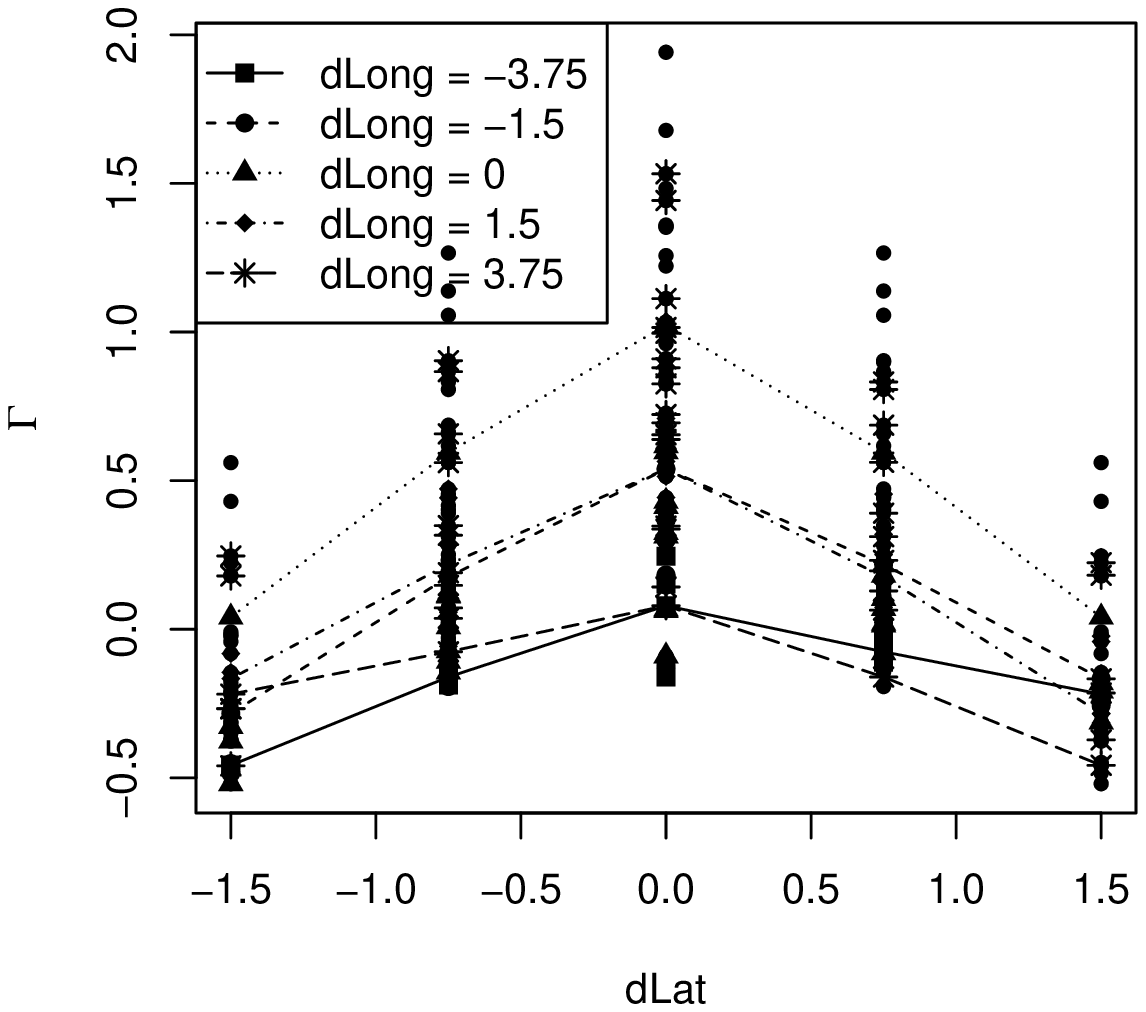}
\vglue.001cm
\includegraphics[width=6.5cm]{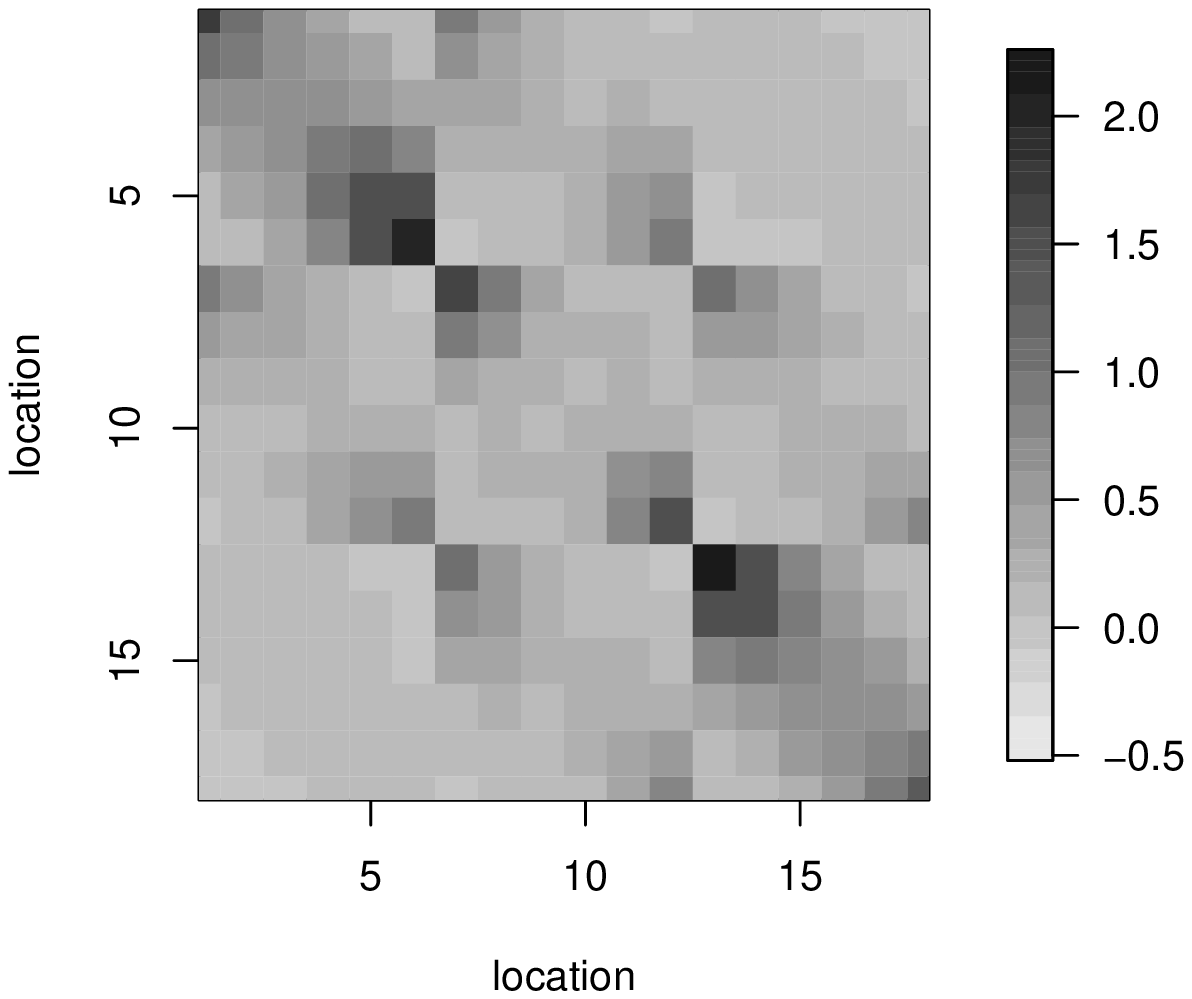}
\hglue.001cm
\includegraphics[width=6.5cm]{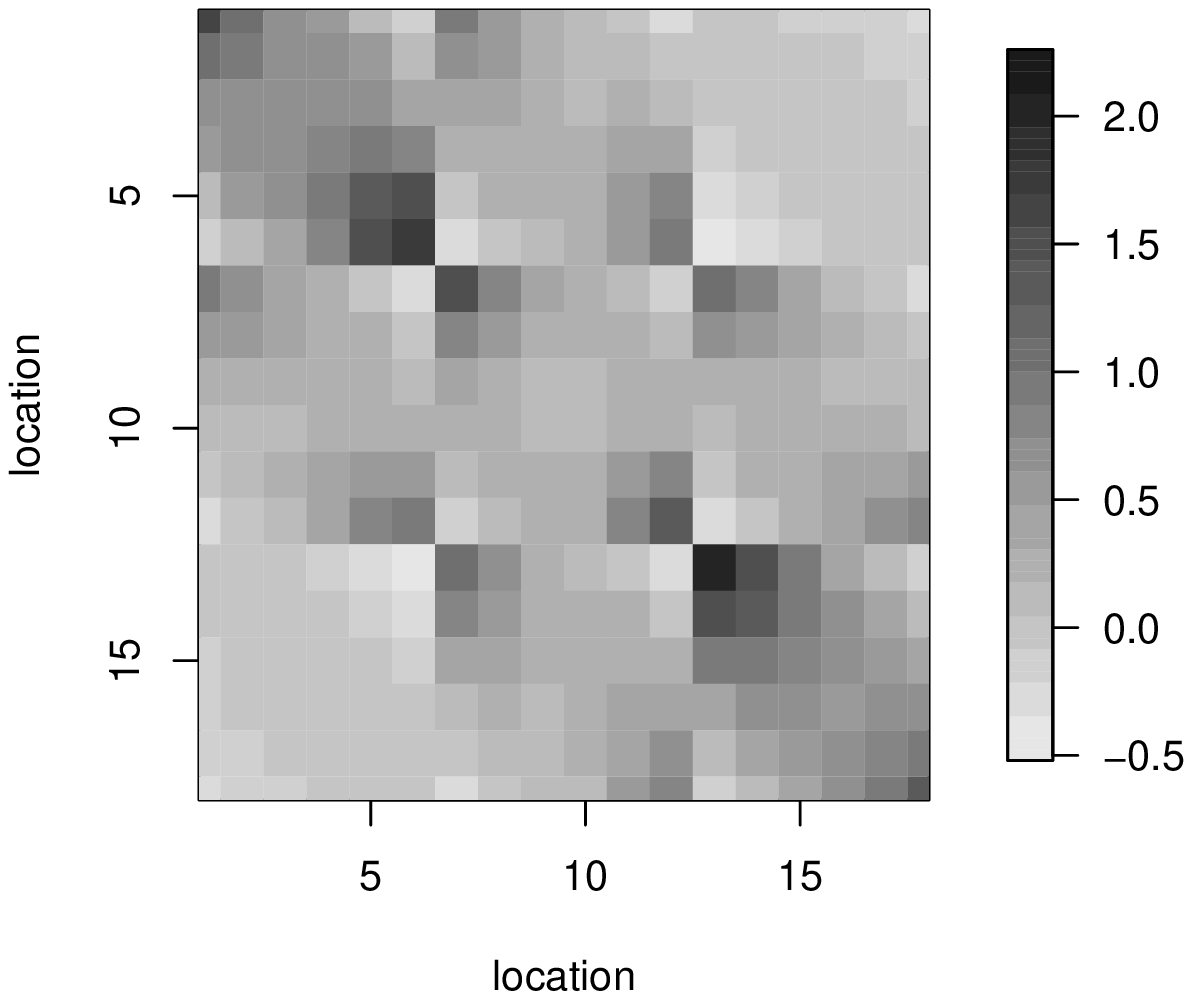}
\caption{Top panels: estimate of $\Gamma$ as a function of the difference in longitude (left) and latitude (right), solid lines: mean values according a given latitude or longitude. Bottom panels: image of covariance matrices fitted by least square to the matrix shown on Figure \ref{imageGamma} (Gaussian covariance (left), wave covariance (right)).}
\label{structR1}
\end{center}
\end{figure}

The spatial structure of the estimated $\Gamma$ shown on Figures \ref{imageGamma} and \ref{structR1} suggests modeling the covariance between locations $i$ and $j$ in $\{1,...,K\}$ as a function of the distance $\mathrm{d}_{{i,j}}$ between these locations. In the sequel we consider two different models, one with Gaussian correlation function 
$$\Gamma_{i,j} =\sigma_{i} \sigma_{j} ( \exp(-\lambda_{1} \mathrm{d}_{i,j}^{2}) + \lambda_{2} \delta_{i,j}) \textrm{ for $i,j$ $\in\{1,...,K\}$,} $$
and the other with wave correlation function 
$$\Gamma_{i,j} =\sigma_{i} \sigma_{j} \Big(  \frac{\sin(\lambda_{1} \mathrm{d}_{i,j})}{\lambda_{1} \mathrm{d}_{i,j}} + \lambda_{2} \delta_{i,j} \Big)\textrm{ for $i,j$ $\in\{1,...,K\}$,} $$
where $(\sigma_{1},...,\sigma_{K},\lambda_{1},\lambda_{2})$ are positive parameters and $\delta_{i,j}$ denotes the Kronecker delta. $\lambda_{1}$ and $\lambda_{2}$ model respectively the range and nugget parameters, and $\sigma_{i} (1+\lambda_{2})$ represents the standard deviation of the field at location $i$. These models are usual well defined covariance functions (see e.g. \cite{CRES91,ABRA97}). They will be denoted respectively ($\M_{\Gamma \sim Gauss}$) and ($\M_{\Gamma \sim Sinus}$) in the sequel. 

The difference in dependence from latitude and longitude of $\Gamma$ (Figure \ref{structR1}) suggests the use of an anisotropic distance (see \cite{REF11,HAS07,SAL11})
 $$\mathrm{d}_{i,j} = \sqrt{\Delta \lat(i,j)^2 + \theta_{1}\Delta \lgtd(i,j)^2 + \theta_{2} \Delta \lat(i,j) \Delta \lgtd(i,j)}$$ 
where $\Delta \lat(i,j)$ and $\Delta \lgtd(i,j)$ denote respectively the difference in latitude and longitude between locations $i$ and $j$ expressed in kilometers. The constraint  $ \theta_{1} >  \frac{\theta_{2}^{2}}{4}$ is imposed to ensure positive-definiteness of the distance. 


These covariance structures have first been fitted by least square estimation to the estimated $\Gamma$ shown on Figure \ref{imageGamma} and the results are shown on Figure \ref{structR1}. The fit is globally satisfying for the wave covariance whereas the Gaussian shape can not cope with the negative correlations observed between western and eastern locations. However the covariance between the northern and southern locations are poorly reproduced (bottom left corner and top right corner of the images of $\Gamma$ in Figures \ref{structR1} and \ref{imageGamma}). As mentioned in Section \ref{sec:inter} these blocks have a particular elliptical shape which can not be reproduced by the parametric models.

Anisotropy coefficients are for the sinus structure: $(\theta_{1},\theta_{2})=(0.2,0.04)$ and for the Gaussian covariance $(\theta_{1},\theta_{2})=(0.23,0.005)$. $\theta_{1} \leq1$ which is reasonable as  the spatial range of the coefficients of $\Gamma$ in longitude is weaker than the one in latitude (Figure \ref{structR1}). The interaction between latitude and longitude is very weak and almost non existing for the Gaussian shape.

\begin{figure}
\begin{center}
\includegraphics[width=6.5cm]{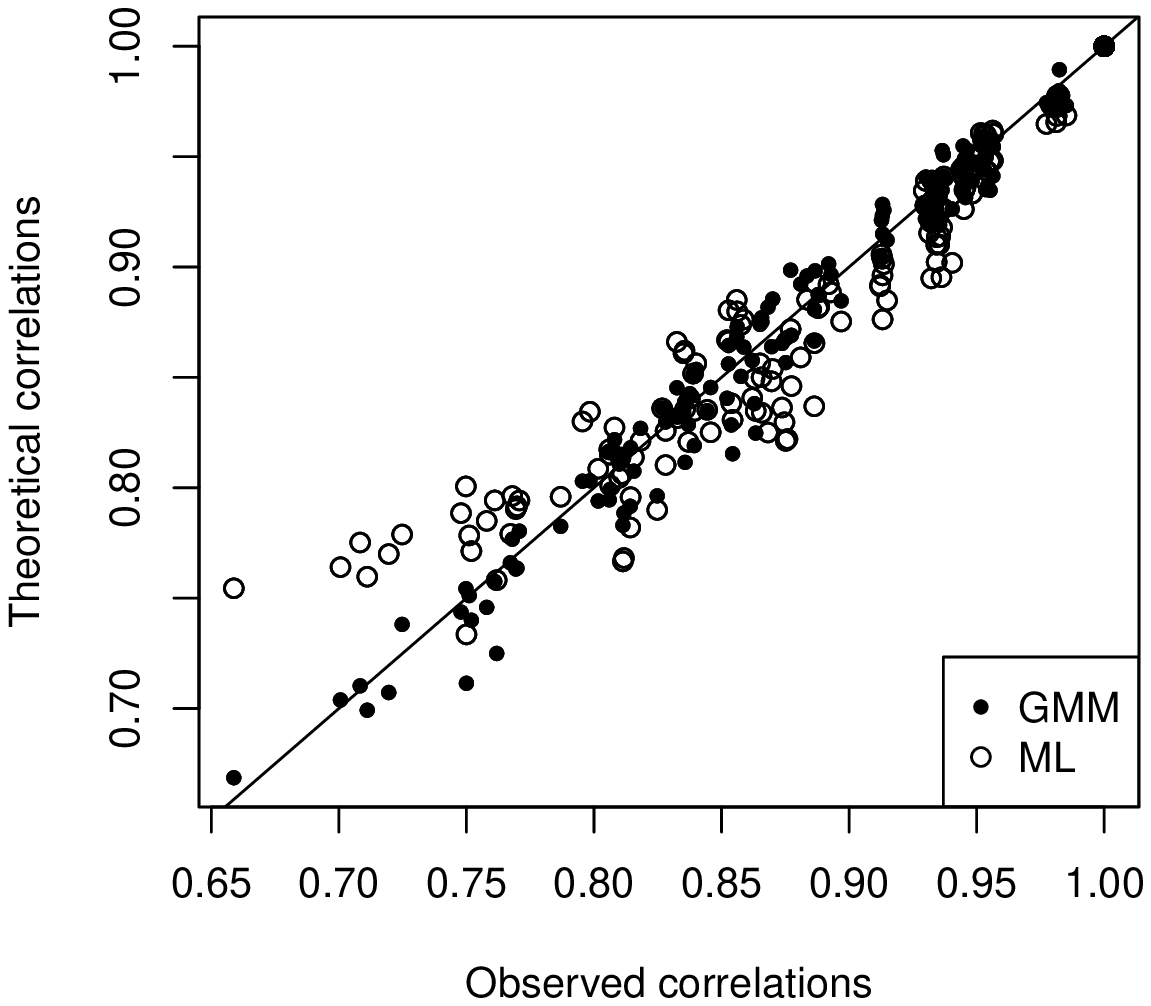}
\hglue.0001cm
\includegraphics[width=6.5cm]{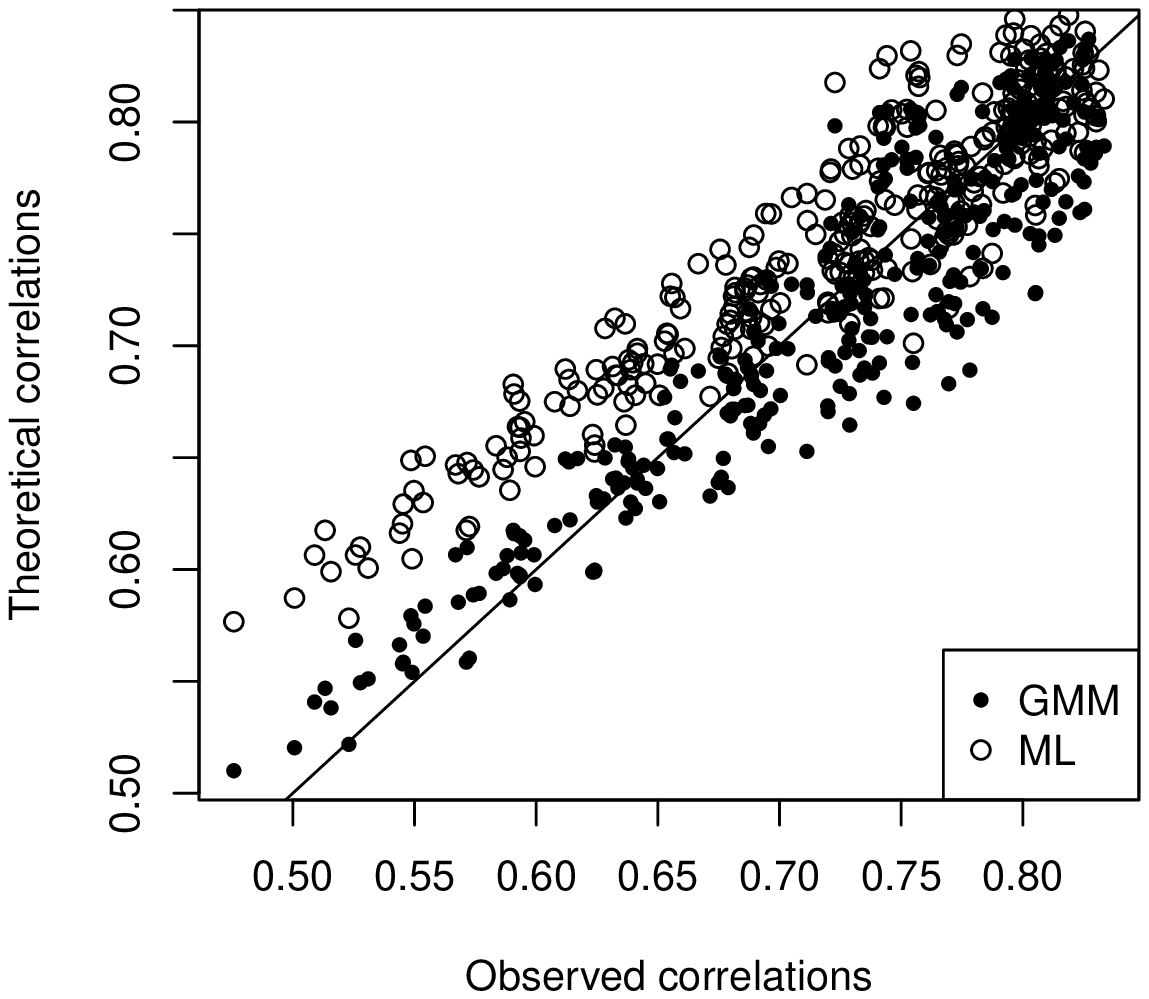} 
\caption{Theoretical correlations of the process $Y$ for model ($\M_{\Gamma \sim Sinus}$) against observed ones at lag $0$ (left) and lag $1$ (right).}
\label{structR2}
\end{center}
\end{figure}

In a second step, the parameters have been re-estimated using the GMM and ML methods. A numerical optimization needs to be performed in the M-step of the EM algorithm to update the values of $(\sigma_{1},...,\sigma_{K},\lambda_{1},\lambda_{2})$. Note that the function to minimize can be expressed in a compact way (see supplementary materials) which leads to an efficient numerical procedure. 
The models have been validated in the same way than model ($\M$) (see Section \ref{results}).  Similar results were obtained as concerns the marginal distributions and the temporal correlation functions but the description of the spatial structure is deteriorated when using a ($\M_{\Gamma}$) model instead of ($\M$) (see Figure \ref{structR2}). This miss-specification is also confirmed by  the Bayes Information Criterion (BIC) and MSPE values given in Table \ref{BIC}  where   $\mathrm{BIC}=-2\log \mathrm{L} + \N_{p} \log(\N_{obs})$ with L the likelihood of the model, $\N_{p}$ the number of parameters and $\N_{obs}$ the number of observations. 
The reduced models ($\M_{\Gamma}$) is clearly outperformed by the full model ($\M$). Other parametric models such as the Mat\'ern family have been tried without more success and it seems difficult to find a simple reduced model which can reproduce all the complexity of the observation error $\Gamma$.

\subsection{Parameterization of $\Lambda$}\label{paramLambda}

\begin{figure}
\begin{center}
\includegraphics[width=6.5cm,angle=90]{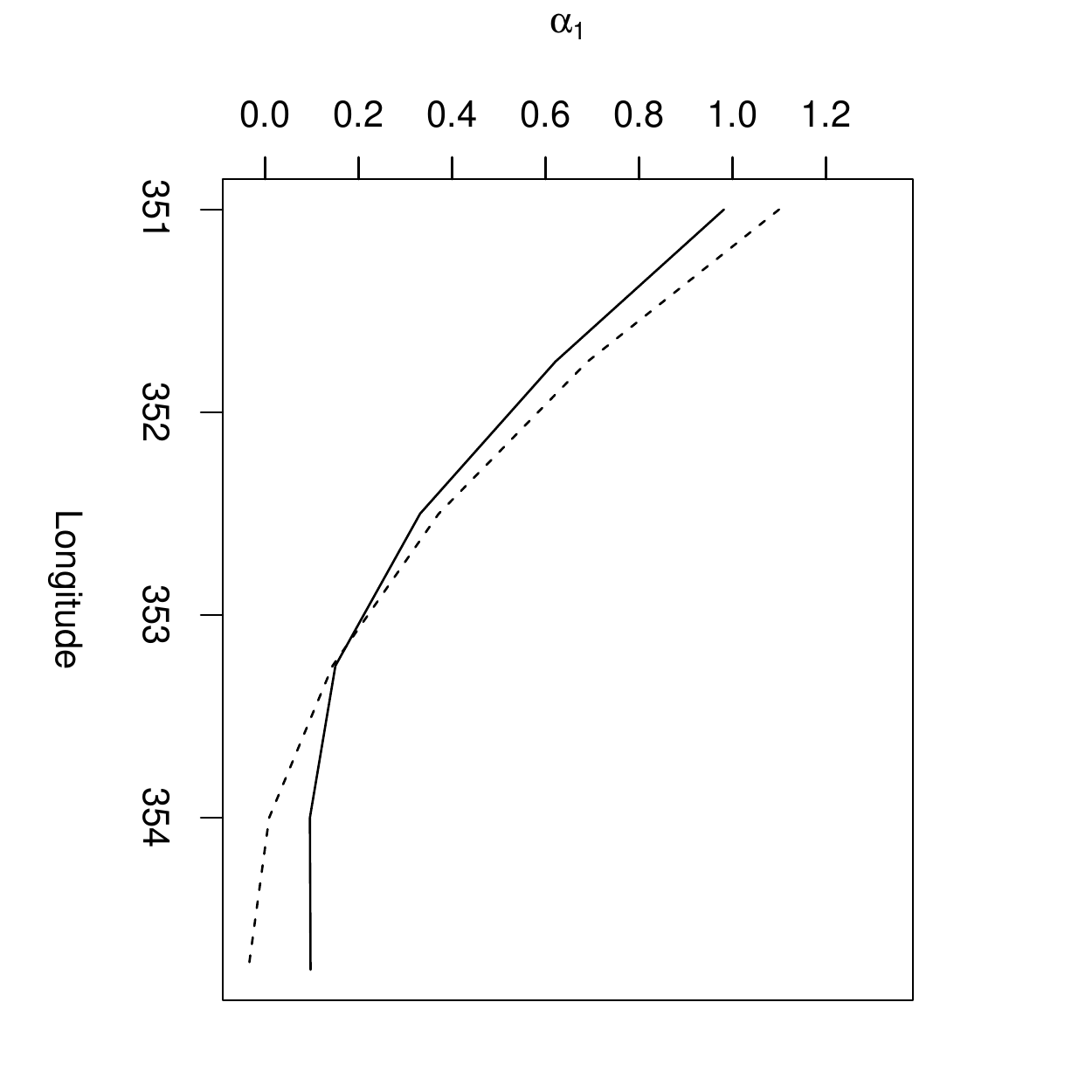}
\hglue.001cm
\includegraphics[width=6.5cm]{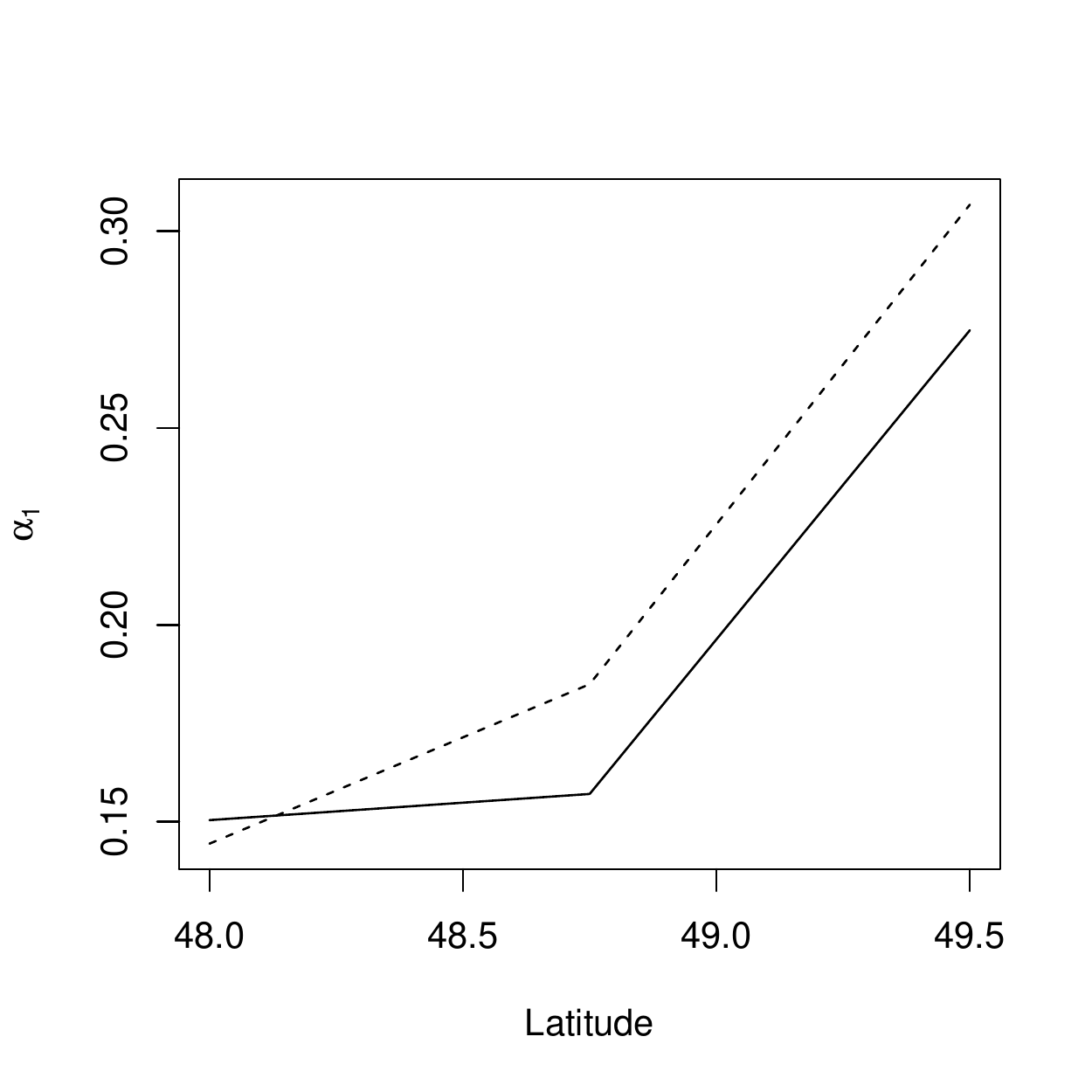}
\vglue.001cm
\includegraphics[width=6.5cm]{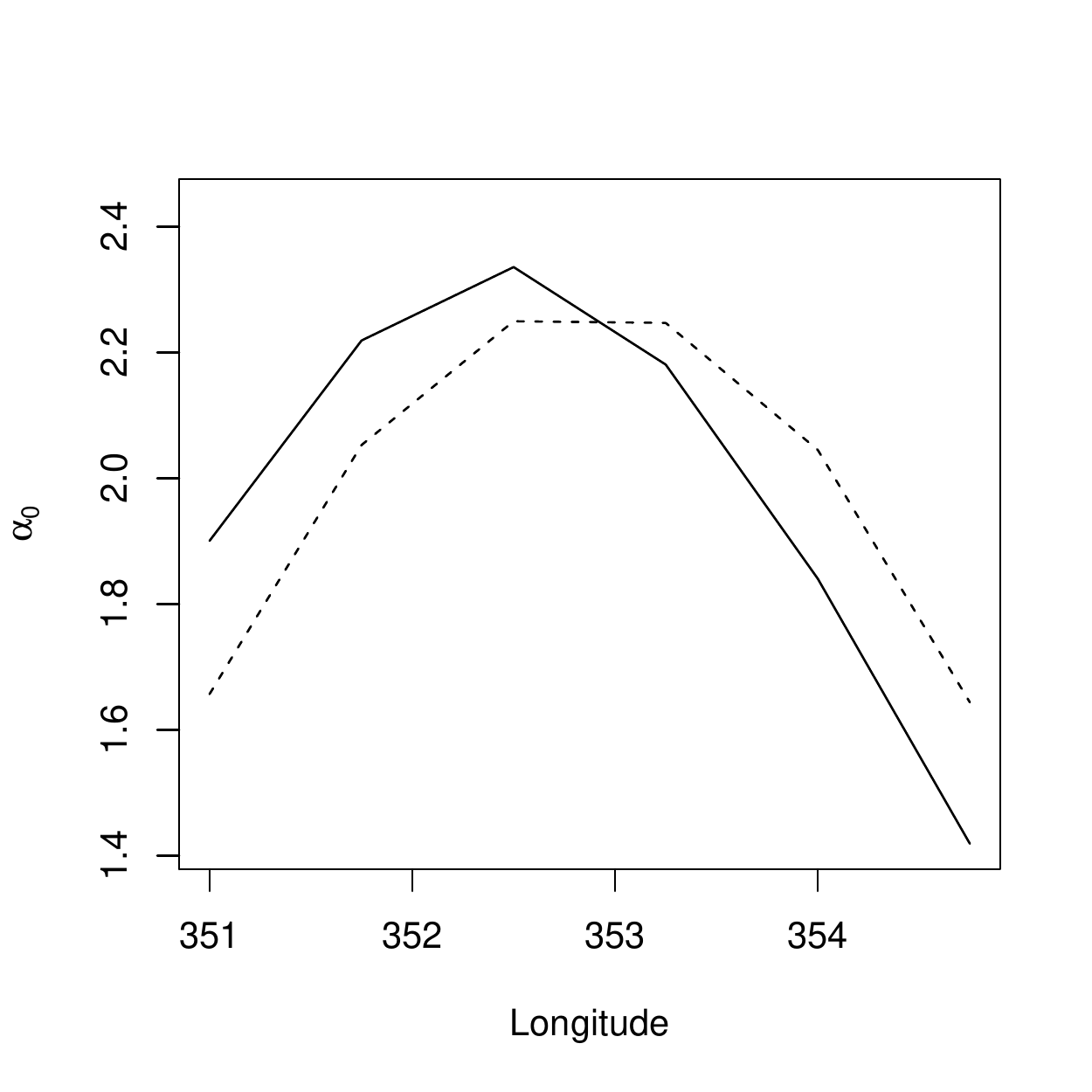}
\hglue.001cm
\includegraphics[width=6.5cm]{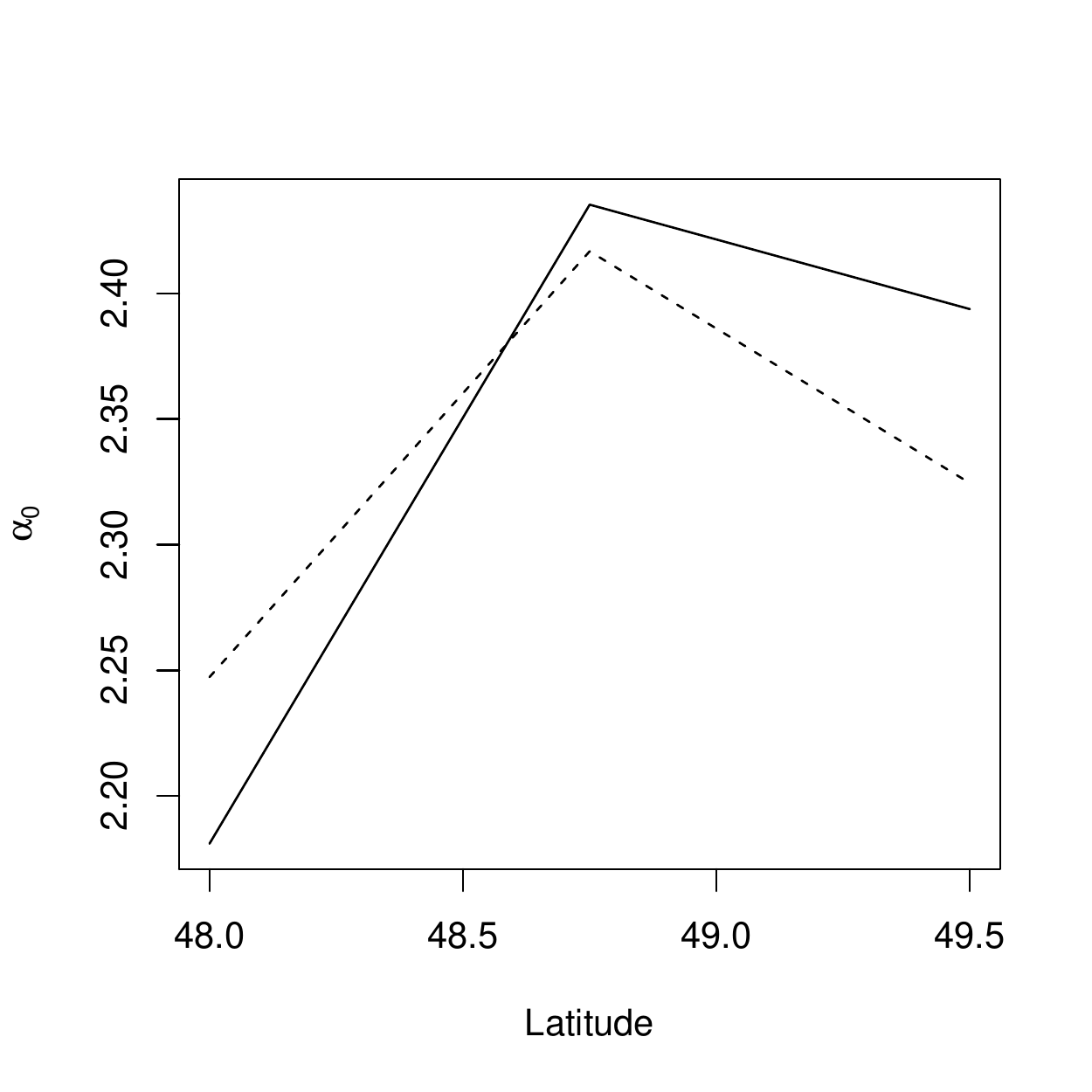}
\vglue.001cm
\includegraphics[width=6.5cm,angle=90]{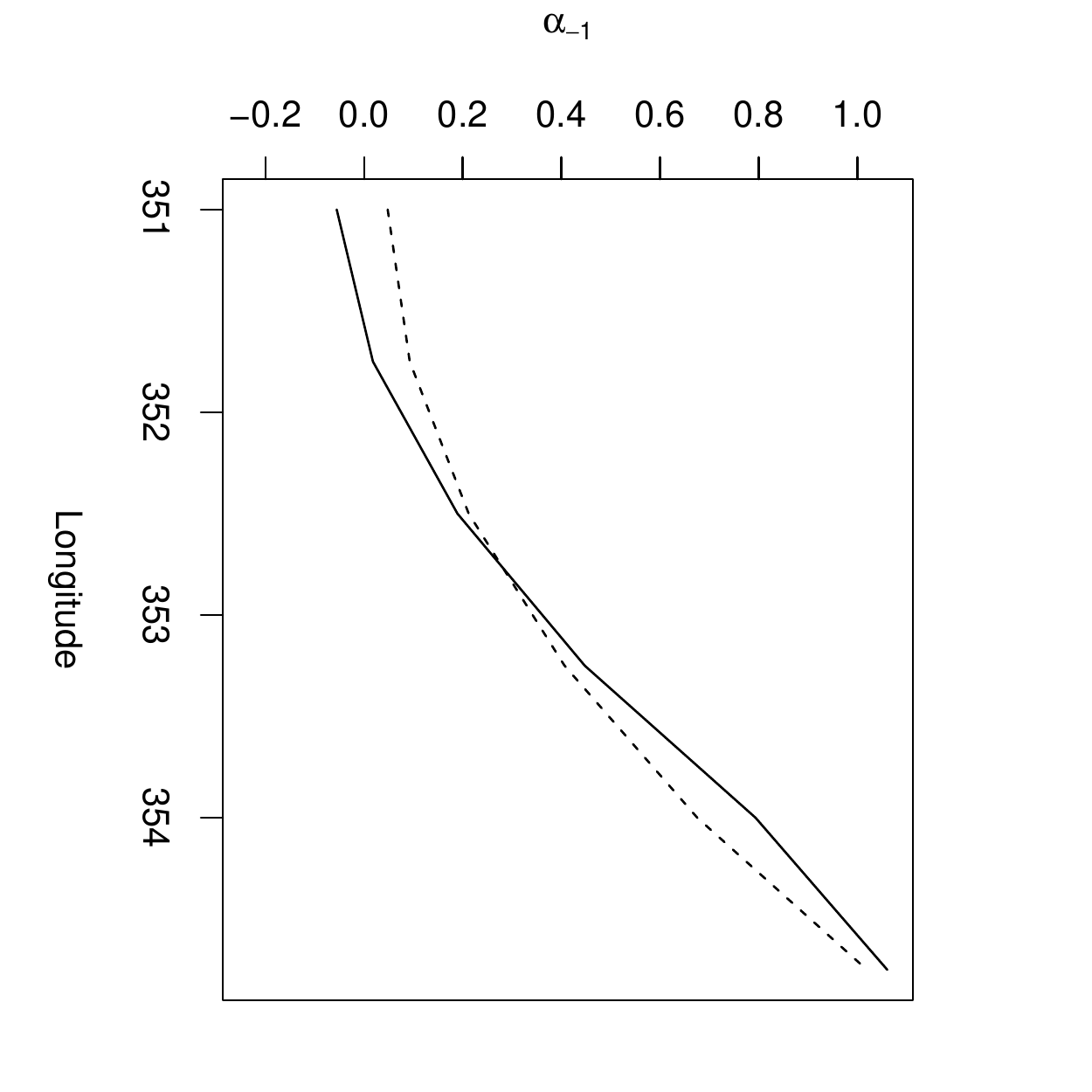}
\hglue.001cm
\includegraphics[width=6.5cm]{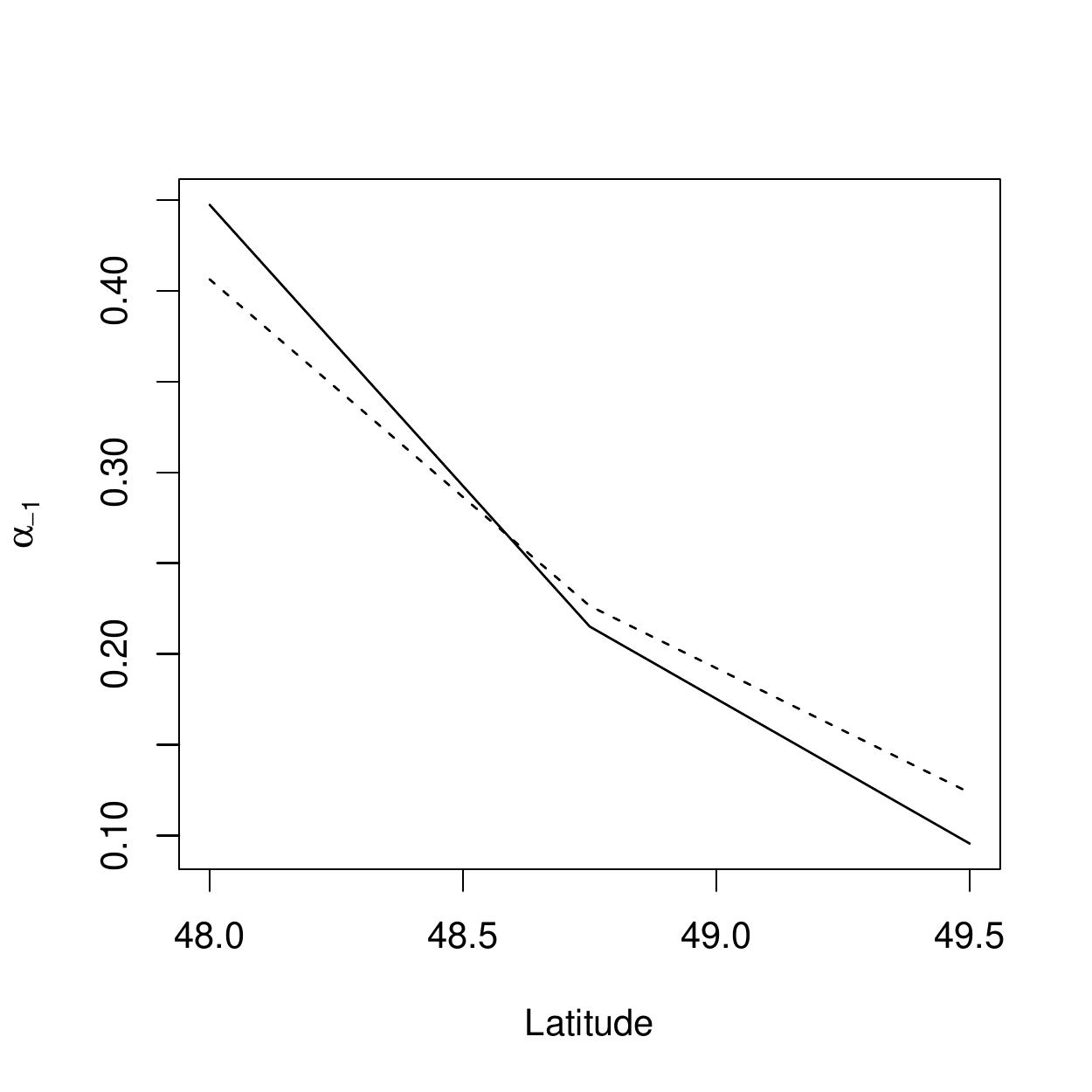}
\caption{Estimated $\alpha_{1}$ (top), $\alpha_{0}$ (middle) and $\alpha_{-1}$ (bottom) against longitude at latitude $48^{\circ}$ N (left) and against latitude at longitude $6.75^{\circ}$ W (right). Solid line: ML estimation of $\Lambda$ for model ($\M$) , dashed line: parametric structure fitted by least square.
}
\label{structZ1}
\end{center}
\end{figure}

The structure of $\alpha_{1}$, $\alpha_{0}$ and $\alpha_{-1}$ reveals a quadratic dependence in longitude and the dependence in latitude suggests the use of an intercept depending on latitude (see Figure \ref{structZ1}). 
This  following parameterization is then proposed.
$$\Lambda = \left( \begin{array}{ccccc} 1 & | & \lgtd  & | & \lgtd^{2} \end{array} \right) \left( \begin{array}{ccc}
							\beta_{1}^{\lat} & \beta_{4}^{\lat} & \beta_{7}^{\lat} \\
							\beta_{2} & \beta_{5} & \beta_{8}  \\
							\beta_{3} & \beta_{6} & \beta_{9} \end{array} \right)$$ 
where $\beta_{i}^{\lat}$ for $i \in\{1,4,7\}$ takes a different value for each latitude and $\lgtd \in \mathbb R^K$ is a vector containing the longitude of the sites. Let ($\M_{\Lambda}$) denote the corresponding model. 
$\Lambda$ is of rank $3$ if the matrix $\left( \begin{array}{ccc}
							\beta_{1}^{\lat} & \beta_{4}^{\lat} & \beta_{7}^{\lat} \\
							\beta_{2} & \beta_{5} & \beta_{8}  \\
							\beta_{3} & \beta_{6} & \beta_{9} \end{array} \right)$ is full ranked because the matrix $\left( \begin{array}{ccccc} 1 & | & \lgtd  & | &  \lgtd^{2} \end{array} \right)$ is full ranked.
							
\begin{figure}
\begin{center}
\includegraphics[width=6.5cm]{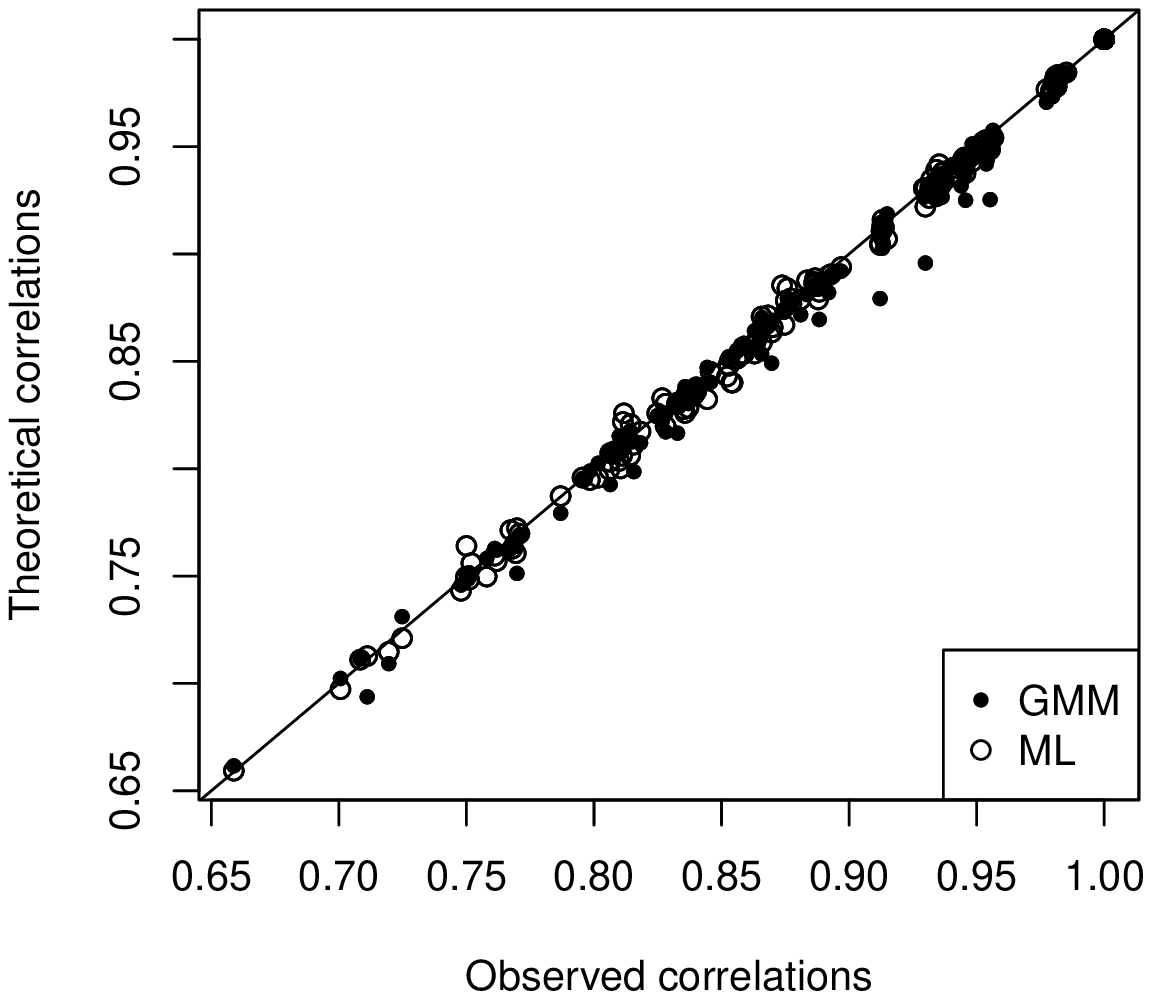} 
\hglue.0001cm
\includegraphics[width=6.5cm,angle=90]{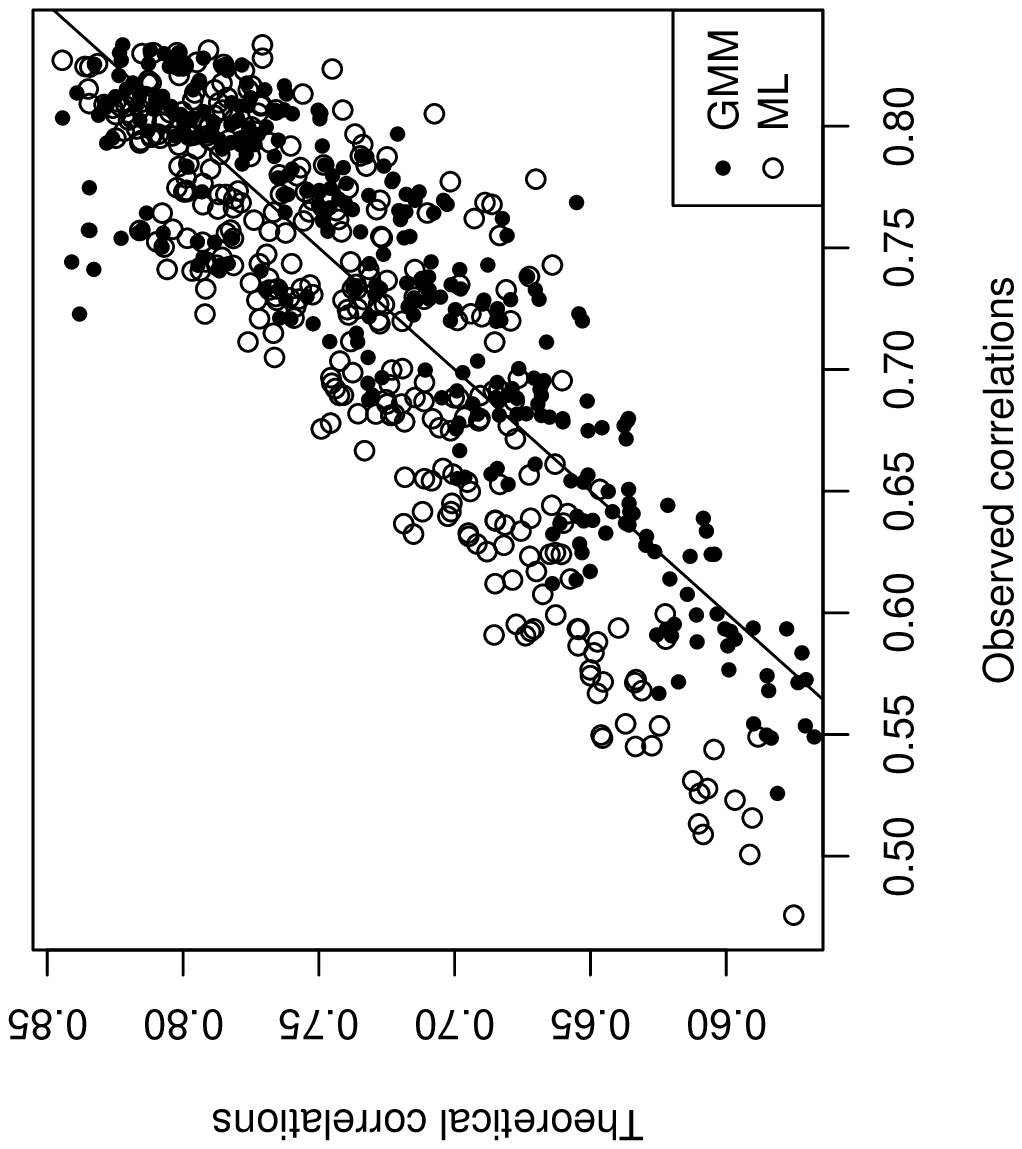} 
\caption{Theoretical correlations against observed ones at lag $0$ (left) and lag $1$ (right) for the model $(\M_{\Lambda})$. }
\label{structZ2}
\end{center}
\end{figure}

The parameterization is easily handled  in the GMM procedure whereas a numerical optimization is again needed to update $\Lambda$ in the M-step. Moreover a joint optimization on $\Lambda$ and $\Gamma$ should be done since both of them are involved in the same part of log-likelihood. In order to avoid a numerical optimization in a high dimensional space, separate optimizations in $\Lambda$ and in $\Gamma$ have been performed leading to a so-called Generalized EM algorithm (see the supplementary materials for more details). The reduced ($\M_{\Lambda}$) and the full ($\M$) models give again similar results for the marginal distribution and the autocorrelation function. ($\M_{\Lambda}$) also leads to an accurate description of the spatial structure of the data (see Figure \ref{structZ2}). Lagged one correlations are better reproduced by GMM parameters than ML parameters. The model ($\M_{\Lambda}$) is slightly inferior to the full model ($\M$) according to the BIC and MSPE values given in Table \ref{BIC} but clearly outperforms the models ($\M_{\Gamma}$). It seems easier to find an appropriate reduced model for the loading matrix $\Lambda$ than for the covariance matrix of the observation error $\Gamma$.

\subsection{The hidden state as an AR(2) process}
\label{AR2}

Higher order autoregressive models have been considered for modeling the dynamics of the hidden state. The best results have been obtained with an AR(2) model defined as 
$$
( \M_{2})\left \{
 \begin{array}{lcl}
X_{t+1}  &=& \rho_{1} X_{t} + \rho_{2} X_{t-1}  + \sigma\epsilon_{t+1},\\
Y_{t} &=& \alpha_{1} X_{t+1} + \alpha_{0} X_{t} + \alpha_{-1} X_{t-1} + \Gamma^{-1/2} \eta_{t}
 \end{array}
\right. \textrm{for $t\geq 0$.}$$
($\M_{2}$) has been fitted using the same procedure than  for model ($\M$) and ($\M_{2}$). The ML estimate of $\rho_{1}$ and $\rho_{2}$ are respectively $0.91$ and $-0.11$. They are close to the values obtained when fitting an AR(2) process to wind data at a single location (see \cite{AILL12}). ($\M_{2}$) slightly outperforms ($\M$) according to the criteria considered in Section \ref{results} and the values given in Table \ref{BIC}. According to Figure~\ref{rmse}, the gain of using the ($\M_{2}$) model to produce short term forecasts is more important at central locations where ($\M_{2}$) and VAR(1) give similar results.

\begin{table}
\begin{center}
\scalebox{.7}{
\begin{tabular}{|c|c|c|c|c|c|}
\hline
Model & Parameters & Log-likelihood & BIC & \multicolumn{2}{c|}{MSPE [min ; max]}\\
 & & & & GMM & ML \\
\hline
VAR(1) & 495 & -20707 & 46961 &  & [ 0.249 ; 0.350 ] \\
\hline
($\M_{2}$) & 209 & -24849 & 52040 &  [ 0.268 ; 0.410 ] & [ 0.256 ; 0.418 ] \\
\hline
($\M$) & 208 & -24954 & 52238 &  [ 0.264 ; 0.410 ] &  [ 0.264 ; 0.414 ] \\
\hline
($\M_{\Lambda}$)  & 186 & -25399 & 52895 &  [ 0.277 ; 0.428 ] &  [ 0.264 ; 0.417 ] \\
\hline
($\M_{\Gamma \sim Gauss}$) & 78 & -29110 & 59082 &  [ 0.308 ; 0.428 ] &  [ 0.274 ; 0.389 ] \\
\hline
($\M_{\Gamma \sim Sinus}$)& 78 & -35615 & 72094 &  [ 0.349 ; 0.478 ] &  [ 0.292 ; 0.403 ] \\
\hline
ARMA(2,1) & 72 &  &  &  &  [ 0.366 ; 0.405 ] \\
\hline
Persistence forecast &  &  &  &  &  [ 0.423 ; 0.468 ] \\
\hline
\end{tabular}}
\caption{\label{BIC} Table of log-likelihoods and BIC indexes for the different models and Mean Square Percentage Error of one-step ahead forecasts by these models.}
\end{center}
\end{table}


\section{General discussion and perspectives}\label{discussion}

Several multisite models, all based on Gaussian linear state-space models, are proposed for wind speed. The main innovation with respect to the other space-time models which have been proposed for meteorological variables is the introduction of a latent process which describes regional conditions. It leads to interpretable models which can reproduce the marginal distribution of wind speed and important properties of the space-time covariance structure such as the asymmetries induced by prevailing motions of the air masses.

An important advantage of Gaussian linear state-space models is that efficient and easy to implement procedure can be used to fit the model. Two estimation procedures, one based on a method of moment (GMM) and the other on the likelihood function (ML) have been compared. GMM appears to be better when looking at the short-term space-time structure but ML seems to better capture  the long-term dynamics. According to Table \ref{BIC}, ML estimates also generally lead to more accurate short-term forecasts except for the model ($\M$). With this latter the southern line (points numbered 1 to 6) is better predicted by GMM parameters than by ML parameters whereas for the other models prediction is better with ML parameters at almost all stations.

According to Table \ref{BIC} the rankings of the models according to BIC or MSPE, which measures the accuracy of one-step ahead forecasts on a validation set, are generally consistent. It indicates that BIC can be trusted when selecting the best model despite the very high conditioning number (ratio of the greatest eigenvalue over the smallest one) of the estimated matrix $\Gamma$ which inverse appears in the log-likelihood function. The ranking also coincides with the complexity of the model: the quality of the model is deteriorated when the number of parameters is reduced. It highlights the difficulty to find parsimonious and realistic models for describing the space-time evolution of the wind. Similar results have been obtained on the Irish wind dataset considered in \cite{RAFT89,GNEI02} which has a different space-time sampling with daily data and stations on an irregular spatial grid.  

The proposed models could be improved in several ways. In particular, we would like to introduce covariates such as large-scale climate variables in order to better understand the interannual variability (see \cite{AILL12}) or other meteorological variables  which could help improving the quality of the short term forecasts (e.g. wind direction, pressure). Alternatively we could include a discrete component within the latent process (a hierarchical model) to describe the regime shifts induced by the weather types (e.g. cycloning conditions with westerly flows and anticycloning conditions with easterly flows).

\appendix{
\section{Proof of proposition \ref{propid}}
\label{nonidentif}

Let $\{Y_t\}$ [resp. $\{\tilde Y_t\}$] denote a process satisfying ($\M$) with parameters $\theta=(\rho,\sigma,\Lambda,\Gamma)$ [resp. $\tilde \theta =(\tilde \rho,\tilde \sigma,\tilde \Lambda,\tilde \Gamma)$]. We assume that $\frac{\sigma^2}{1-\rho^2}=1$ and $\Lambda$ is full ranked, with the same constraints holding true for $\tilde \theta$. We also assume that $\{Y_t\}$ and $\{\tilde Y_t\}$ have the same second order structure. We prove below that if these conditions hold true then $\theta = \tilde \theta$ up to the sign of $\Lambda$ \textit{i.e.} $\rho=\tilde \rho$, $\sigma = \tilde \sigma$, $\Lambda = \pm \tilde \Lambda$ and $\Gamma=\tilde \Gamma$. The proof is based on the properties of $C_k=\cov(Y_t,Y_{t+k})$.

\begin{itemize}
	\item \textbf{Identification of $\rho$ and $\sigma$.} According to (\ref{cov2}), we have $C_k = \rho^{k-2}C_2$ for $k \geq 2$ and
	$$C_2=\frac{\sigma^2}{1-\rho^2} u v^{t}$$ 
	with $u = \alpha_{1} + \rho \alpha_{0} + \rho^{2} \alpha_{-1}$ and  $v= \rho^{2}\alpha_{1} + \rho \alpha_{0} + \alpha_{-1}$. Since $\alpha_{-1}$, $\alpha_{0}$ and $\alpha_{1}$ are linearly independent, $u\neq 0$ and $v \neq 0$ and thus $C_2 \neq 0$. $\rho$ can thus be expressed as a ratio between some coefficients of $C_3$ and $C_2$ and we deduce that $\rho=\tilde \rho$. Using the constraint $\frac{\sigma^2}{1-\rho^2}=1$, we also deduce that $\sigma^2=\tilde \sigma^2$.
	\item \textbf{Identification of $\Lambda$ when $\rho \neq 0$.}  According to (\ref{cov1}-\ref{cov2}) we have $C_2-\rho C_1 = (1-\rho^2)\alpha_{1} \alpha_{-1}^{t}$ and thus $\alpha_{1} \alpha_{-1}^{t}=\tilde \alpha_{1} \tilde \alpha_{-1}^{t}$ since $\rho^2 \neq 1$. We deduce that there exists a real constant $k_1 \neq 0$ such that 
	$\alpha_{-1}=k_1 \tilde \alpha_{-1}$ and $\alpha_{1}=k_1^{-1} \tilde \alpha_{1}$. We also have $u v^{t} = \tilde u \tilde v^{t}$ where $\tilde u$ and $\tilde v$ are defined similarly to $u$ and $v$. We deduce that there exists a real constant $k_2 \neq 0$ such that $\tilde u=k_2 u$ and $\tilde v=k_2^{-1} v$ and thus $\tilde u-\tilde v= k_2 u -k_2^{-1} v$ with
	\begin{eqnarray}
	\nonumber
	\tilde u-\tilde v &=&(1-\rho^2) \tilde \alpha_1 + (\rho^2-1)\tilde \alpha_{-1}\\
	\label{tilde}
	&=&(1-\rho^2) k_1^{-1} \alpha_1 + (\rho^2-1) k_1 \alpha_{-1}\\
	\label{notidle}
	k_2 u -k_2^{-1} v&=&(k_2-\rho^2 k_2^{-1})\alpha_1+\rho(k_2-k_2^{-1})\alpha_0 + (k_2\rho^2 -k_2^{-1})\alpha_{-1}  
	\end{eqnarray}
Since $\alpha_{-1}$, $\alpha_{0}$ and $\alpha_{1}$ are linearly independent, we can identify the coefficients of the linear combinations (\ref{tilde}-\ref{notidle}) and deduce, when $\rho \neq 0$ that  $k_2 \in \{-1,1\}$ and $\alpha_i = k_2 \tilde \alpha_{i}$ for $i \in \{-1,0,1\}$. 
\item \textbf{Identification of $\Lambda$ when $\rho = 0$.} In this case, 
\begin{eqnarray}
 \label{c1}
 C_{1} & = &  \sigma^{2}(\alpha_{1} \alpha_{0}^{t} + \alpha_{0} \alpha_{-1}^{t}), \\
  \label{c2}
 C_{2} & = & \sigma^{2}\alpha_{1} \alpha_{-1}^{t}
 \end{eqnarray}
By similar reasoning as previously from (\ref{c2}) there exists $k_{1}\neq 0$ such that $\alpha_{-1}=k_1 \tilde \alpha_{-1}$ and $\alpha_{1}=k_1^{-1} \tilde \alpha_{1}$. From (\ref{c1})  we deduce that $\alpha_{1}(k_{1}\tilde \alpha_{0} - \alpha_{0})^{t} +  (\frac{\tilde{\alpha_{0}}}{k_{1}} - \alpha_{0}) \alpha_{-1}^{t} = 0$. 
  
  If $k_{1}\tilde \alpha_{0} - \alpha_{0} \neq 0$ then there exists $k_{2}\neq 0$ such that $\alpha_{1} - \frac{k_{2}}{k_{1}} \tilde \alpha_{0} - k_{2} \alpha_{0}=0$ $(\R_{1})$ and $ \frac{1}{k_{2}} \alpha_{-1} + \alpha_{0} + k_{1}\tilde \alpha_{0} = 0$ $(\R_{2})$. Then $$ (\R_{1}) - \frac{k_{2}}{k_{1}} (\R_{2}) = \alpha_{1} + (k_{2} + \frac{k_{2}}{k_{1}^{2}})\alpha_{0} + \frac{1}{k_{1}^{2}}\alpha_{-1} = 0.$$ Since $\alpha_{1}$, $\alpha_{0}$ and $\alpha_{-1}$ are linearly independent we obtain $k_{1}=k_{2}=0$ which is a contradiction.

  If $k_{1}\tilde \alpha_{0} - \alpha_{0} = 0$, this implies $\frac{\tilde{\alpha_{0}}}{k_{1}} - \alpha_{0} = 0$, then $k_{1} = \pm 1$. In both cases, $\alpha_{1}$, $\alpha_{0}$ and then identifiable from the covariance $C_{2}$ and $C_{1}$.

\item \textbf{Identification of $\Gamma$.} According to (\ref{cov0}), $\Gamma$ can be expressed from $C_0$ and the other parameters. We easily deduce that $\tilde \Gamma = \Gamma$
\end{itemize}

 Here we prove that full-symmetry can not be achieved under the chosen identifiability constraints. Separability of a space-time covariance function implies full-symmetry of this latter (\cite{GNEI02}). Full-symmetry of the space-time covariance function implies that the matrix $C_{2}$ is a symmetric matrix. 
The symmetry of $C_{2}$ implies $u v^{t} = v u^{t}$, $u$ and $v$ are then collinear vectors which implies a collinearity between $\alpha_{1}$, $\alpha_{0}$ and $\alpha_{-1}$. The space-time covariance function defined by the model is not fully-symmetric and then non-separable.}

\section{Supplementary materials}
\begin{description}
\item[MLE for the model (M) and associated reduced models:] This file contains a description of the Expectation-Maximization (EM) algorithm used to fit the model (M) and the associated reduced models (supp$\_$estimation.pdf).
\end{description}

\bibliographystyle{alpha}
\bibliography{biblio_articleGSSM}

\end{document}